\documentclass[12pt,reqno]{amsart}

\usepackage{epsfig}
\usepackage{amsmath, amsthm, amsfonts}
\usepackage{graphicx}

\numberwithin{equation}{section}

\newcommand{\R}{{\mathbb R}}
\newcommand{\C}{{\mathbb C}}

\newcommand{\Z}{{\mathbb Z}}

\renewcommand{\Re}{{\operatorname{Re\,}}}
\renewcommand{\Im}{{\operatorname{Im\,}}}

\newcommand{\dist}{{\operatorname{dist}}}

\newcommand{\Erfc}{{\operatorname{erfc}\,}}

\newcommand{\al}{\alpha}
\newcommand{\be}{\beta}
\newcommand{\ga}{\gamma}
\newcommand{\Ga}{\Gamma}

\newcommand{\la}{\lambda}
\newcommand{\ep}{\varepsilon}
\newcommand{\de}{\delta}

\newcommand{\f}{\varphi}
\newcommand{\sg}{\sigma}
\newcommand{\om}{\omega}
\newcommand{\Om}{\Omega}
\newcommand{\z}{\zeta}

\newtheorem{theo}{{\sc \bf Theorem}}[section]

\newtheorem{lem}[theo]{{\sc \bf Lemma}}
\newtheorem{prop}[theo]{{\sc \bf Proposition}}

\sc
\title[Zeros of Sections of Exponential Sums]
{Zeros of Sections of Exponential Sums}
\author{Pavel  Bleher}
\address{Department of Mathematical Sciences, Indiana
University-Purdue University Indianapolis,
402 N. Blackford St., Indianapolis, IN 46202, U.S.A.}
\email{bleher@math.iupui.edu}
\author{Robert Mallison, Jr.}
\address{Department of Mathematics, Indiana Wesleyan University,
4201 S. Washington St., Marion, IN 46953, U.S.A.}
\email{bob.mallison@indwes.edu}

\date{\today}

\thanks{The first author was supported in part by
NSF Grant DMS-0354962. }

\rm

\begin{document}

\begin{abstract}
We derive the large $n$ asymptotics of zeros of
sections of a generic exponential sum. We divide all the zeros
of the $n$-th section of the exponential sum into ``genuine zeros'', which approach,
as $n\to\infty$, the zeros of the exponential sum, and
``spurious zeros'', which go to infinity as $n\to\infty$. We show
that the spurious zeros, after scaling down by the factor of $n$,
approach a ``rosette'', a finite collection of curves on the complex plane,
resembling the rosette.
We derive also the large $n$ asymptotics of the ``transitional zeros'',
the intermediate zeros between genuine and spurious ones. 
Our results give an extension to the classical results of Szeg\"o
about the large $n$ asymptotics of zeros of
sections of the exponential, sine, and cosine functions.
\end{abstract}

\maketitle

\section {Introduction}

We will be interested in this paper in the distribution 
of zeros of  sections of exponential sums.  
We consider the exponential sum,
\begin{equation}\label{in1}
f(z)=\sum_{j=1}^M c_je^{\la_j z},
\end{equation}
where $c_j,\;\la_j\in\C$, and its Taylor series,
\begin{equation}\label{in2}
f(z)=\sum_{k=0}^\infty a_k z^k.
\end{equation}
The $n$-th section of $f(z)$ is the finite Taylor series,
\begin{equation}\label{in3}
f_n(z)=\sum_{k=0}^n a_k z^k.
\end{equation}
The problem is to find the distribution of zeros of $f_n$,
$f_n(z_k)=0$,
as $n\to\infty$. This problem was posed and solved for
$f(z)=e^z$ in the classical paper of Szeg\"o \cite{Sze}. 
Szeg\"o proved that as $n\to\infty $,
the rescaled zeros,
\begin{equation}\label{in4}
\z_k=\frac{z_k}{n}\,,
\end{equation}
approach the curve,
\begin{equation}\label{in5}
\Gamma=\{\z:\;|e^{1-\z}\z|=1,\quad |\z|\le 1\},
\end{equation}
on the complex plane, and the limiting distribution of the zeros on 
$\Ga$ is the measure of the maximal entropy, the preimage of the uniform measure
on the circle under the Riemann map. 
Precise asymptotics of the zeros of sections of $e^z$ and the sections themselves
were obtained in the works of Buckholtz \cite{Buc}, Newman and Rivlin \cite{NR},
Carpenter, Varga, and Waldvogel \cite{CVW}, Pritsker and Varga \cite{PV}.
The absence of zeros in some parabolic domains on the complex plane 
was established in the works
of Newman and Rivlin \cite{NR} and Saff and Varga \cite{SV}.
For connections of zeros of sections of $e^z$ to 
the Riemann zeta-function see the works of Conrey and Ghosh \cite{CG}
and Yildirim \cite{Yil}.

Szeg\"o also found the limiting distribution
of the sections zeros for $f(z)=\cos z$ and $f(z)=\sin z$. 
In this case a part of the zeros of $f_n$
approaches the zeros of $f$ as $n\to\infty$, but there is another part of the zeros,
the ``spurious zeros'', which go to infinity as $n\to\infty$. Szeg\"o proved that
as $n\to\infty$ the rescaled spurious zeros approach a limiting curve and
have a limiting distribution on this curve. Close results were obtained 
by Dieudonn\'e \cite{Die}, by a different method. Detailed asymptotics of 
the zeros of
sections of $\cos z$ and $\sin z$ were obtained in the works of Kappert \cite{Kap}
and Varga and Carpenter \cite {VC1}, \cite{VC2}. 
See also the review  papers of Varga \cite{Var}, Ostrovskii \cite{Ost}, and
Zemyan \cite{Zem}. The distribution of zeros of analytic functions is a 
classical area of complex analysis, and many results concerning the
distribution of zeros of analytic functions are discussed in
the monograph of Levin \cite{Lev}. The distribution of sections of
analytic functions of the Mittag-Leffler type is studied in the work
of Edrei, Saff, and Varga \cite{ESV}.

Our main goal in this work
is to  obtain asymptotics of zeros of sections
of exponential sums. First we discuss, in Section 2, the asymptotics of large
zeros of exponential sums themselves. The rest of the paper is devoted to the
asymptotics of zeros of the sections of exponential sums. As an example,
let us consider the exponential sum,
\begin{equation}\label{in6}
\begin{aligned}
f(z)=3 e^{(8+2i)z}&+(-9+12i)e^{(4+7i)z} +(2+i)e^{(-7+4i)z}-5e^{(-6-6i)z}    
   +(6-7i)e^{(1-8i)z}\\
  &+(8-5i)e^{(6-4i)z}+(3-9i)e^{(4+4i)z}+2ie^{(-2-4i)z}.
\end{aligned}
\end{equation}
The zeros of the section of this function for $n=250$ are depicted on Figure 1. The
zeros form a shape resembling a rosette. In this paper we obtain the
large $n$ asymptotics of the zeros of exponential sums, 
 which provides us with explicit equations for different parts of
the rosette.

\begin{center}
\begin{figure}[h]\label{figure1}
\begin{center}
\scalebox{0.5}{\includegraphics{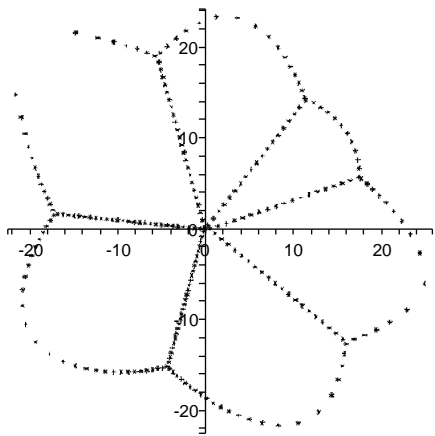}}
\end{center}
   \caption{The zeros of the $n=250$ section of exponential sum (\ref{in6}).}
   \end{figure}
\end{center}

We divide the zeros of $f_n$
into four classes: (1) finite zeros, (2) zeros of the main series,
(3) spurious zeros, and (4) transitional zeros. They are described as follows:
\begin{itemize}
\item
The {\it finite zeros} are the ones that lie in a finite disk,
$D(0,R_0)=\{z\in\C:\; |z|\le R_0\}$. 
\item
The {\it zeros of the main series}
are located in a small neighborhood of the rays, on the intervals $R_0\le |z|\le nr_c(j,n)-R_1$,
$R_1>0$, where $j$ is the number of the ray, and $\lim_{n\to\infty}r_c(j,n)= r_c(j)>0$ is
the {\it critical radius} on the $j$-th ray. We derive 
the angular coordinate of the $j$-th ray and a transcendental equation, which
determines $r_c(j)$ uniquely. 
\item
As $n\to\infty$, both the finite zeros and
the zeros of the main series converge to the zeros of the exponential
sum, $f(z)$. We call them the {\it genuine zeros} of $f_n$.
In addition to them, there are {\it spurious zeros} of $f_n$,
which go to infinity as $n\to\infty$. If we scale down the
spurious zeros by the factor of $n$, they approach to some curves
$\mathcal G_j$. We derive the equations of the curves $\mathcal G_j$.
As a better approximation to the spurious zeros, we construct curves
$\mathcal G_j^n$, which approach $\mathcal G_j$ as $n\to\infty$, and
such that the scaled down spurious zeros lie in the $O(n^{-2})$-neighborhood 
of $\mathcal G_j^n$.
\item
The {\it transitional zeros} of $f_n$ are the intermediate ones, located near the triple points
on Figure 1, where the zeros of the 
main zeries and the spurious zeros
merge. We derive an equation, which gives the asymptotic
location of the transitional zeros. This is determined by zeros of a three term
exponential sum.
\end{itemize}

We derive 
the asymptotics of the zeros of $f_n$, as $n\to\infty$, in Sections 4--9 below.
In Appendices \ref{uniform} and \ref{s_n},
we obtain uniform asymptotics of zeros of the sections of $e^{n\z}$,
and of the sections themselves, in a fixed neighborhood of 
the point $\z=1$. These uniform asymptotics are used in the main
part of the paper to derive the asymptotics of the spurious zeros of $f_n$.

We would like to mention here the work   
of Kuijlaars and McLaughlin \cite{KM}, where the Riemann-Hilbert approach 
to distribution of zeros of Laguerre polynomials with nonclassical parameters is
developed. The distribution
of zeros in \cite{KM} has many similarities to the distribution of
zeros of exponential sums. Also we would like to mention the work 
of Bergkvist and Rullg\aa rd \cite{BR}, in which the distribution of zeros
of polynomial eigenfunctions of some differential equations of higher
order was studied. The distribution of zeros in \cite {BR} seems to have
similarities to the distribution of
zeros of sectios of exponential sums as well.

\section{Zeros of exponential sums}\label{ZES}

We consider the exponential sum,
\begin{equation}\label{es1}
f(z)=\sum_{j=1}^M c_je^{\la_j z},\qquad c_j\not=0,
\end{equation}
where we assume that the numbers $\la_j$ satisfy the following condition:

\vskip 2mm

\noindent
{\bf Condition P.} {\it The numbers
$\la_j$, $j=1,\ldots,m$, $m\ge 3$, are the vertices
of a convex $m$-gon $P_m$ on the complex plane, and  
the numbers $\la_j$, $j=m+1,\ldots,M$, lie
strictly inside of $P_m$.} 

The polygon $P_m$ is the {\it convex hull} of the the numbers $\la_j$, $j=1,\ldots,M$,
on the complex plane, and condition P restricts the remaining
numbers $\la_j$, $j=m+1,\ldots,M$, to lie
strictly inside of $P_m$ (not on the sides of $P_m$).
For the sake of definiteness, we 
will assume that the vertices $\la_1,\ldots,\la_m$ are enumerated
counterclockwise along $P_m$. Figure 2 shows the convex hull for exponential
sum (\ref{in6}).

\begin{center}
\begin{figure}[h]\label{figure2}
\begin{center}
\scalebox{0.5}{\includegraphics{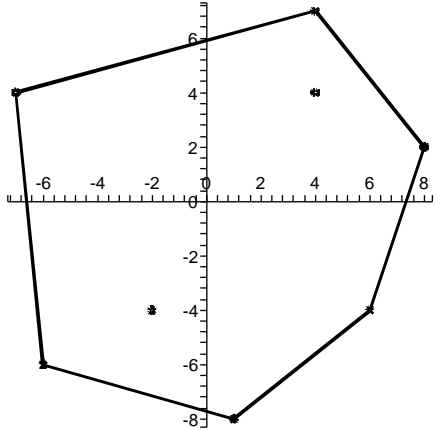}}
\end{center}
   \caption{The convex hull for exponential sum (\ref{in6}), with $\la_1=8+2i,\;
\la_2=4+7i,\;\la_3=-7+4i,\;\la_4=-6-6i,\;\la_5=1-8i,\;\la_6=6-4i,\;\la_7=4+4i,\;\la_8=-2-4i$.}
   \end{figure}
\end{center}

In this section we describe the asymptotics
of zeros of $f(z)$ on the complex plane as $|z|\to\infty$.
We begin with a description of sectors free of large zeros of $f$. 
Define
\begin{equation}\label{es2}
\theta_{jk}=-\arg (\la_{k}-\la_j)+\frac{\pi}{2}\mod 2\pi.
\end{equation}
Partition the complex plane into the sectors,
\begin{equation}\label{es3}
\mathcal U_j=\{z=re^{i\theta}:\;\theta_{j,j+1}
<\theta<\theta_{j-1,j}\mod 2\pi,\;r>0\},
\qquad j=1,\ldots,m,
\end{equation}
where we take the convention that
\begin{equation}\label{es4}
\theta_{m,m+1}=\theta_{01}=\theta_{m1}.
\end{equation}
The notation $\theta_{j,j+1}<\theta<\theta_{j-1,j}\mod 2\pi$ means
that $\theta$ belongs to the interval from $\theta_{j,j+1}$ to
$\theta_{j-1,j}$ on the unit circle in the positive direction.
Define also the rays,
\begin{equation}\label{es5}
\mathcal S_{j,j+1}=\{z=re^{i\theta}:\;\theta=\theta_{j,j+1},\;r\ge 0\},
\qquad j=1,\ldots,m,
\end{equation}
so that $\mathcal U_j$ is the sector between the rays $\mathcal S_{j,j+1}$
and $\mathcal S_{j-1,j}$. Observe that the rays $\mathcal S_{j,j+1}$ are
orthogonal to the sides of the complex conjugate convex hull, $\overline{P_m}$,
of the numbers $\la_j$. Figure 3 shows the complex conjugate convex hull and
the rays $\mathcal S_{j,j+1}$ for exponential sum (\ref{in6}).

\begin{center}
\begin{figure}[h]\label{figure3}
\begin{center}
\scalebox{0.5}{\includegraphics{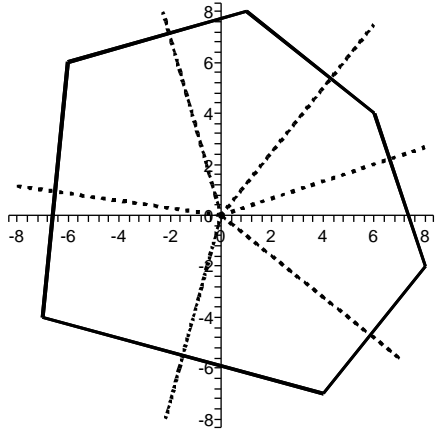}}
\end{center}
   \caption{The complex conjugate convex hull for exponential sum (\ref{in6}),
and the corresponding rays $\mathcal S_{j,j+1}$, $j=1,\ldots,6$.}
   \end{figure}
\end{center}

For a given $j=1,\ldots,m,$ we write
\begin{equation}\label{es6}
\sum_{k=1}^M c_ke^{\la_k z}
=c_je^{\la_j z}\left[1+\sum_{k:\; k\not=j}
\frac{c_k}{c_j}\,e^{(\la_k-\la_j)z}\right].
\end{equation}
We will describe a region where the sum in the brackets on the right is small
and, as a result, $f(z)\not=0$.

\begin{prop} \label{T_j}
Let us fix $\theta$ in the interval
\begin{equation}\label{es7}
\theta_{j,j+1}<\theta<\theta_{j-1,j}\mod 2\pi.
\end{equation}
Then for any $k\not=j$,
\begin{equation}\label{es8}
\lim_{r\to\infty}e^{(\la_k-\la_j)z}= 0,\qquad z=re^{i\theta}.
\end{equation}
\end{prop}

\begin{proof} Since $\la_{j-1},\;\la_j,\;\la_{j+1}$ are the  
vertices of the convex hull of the numbers $\la_k$,
we have that
\begin{equation}\label{es9}
\arg (\la_{j+1}-\la_j)\le \arg (\la_k-\la_j)\le \arg (\la_{j-1}-\la_j)\mod 2\pi,
\end{equation}
hence
\begin{equation}\label{es10}
-\theta_{j,j+1}+\frac{\pi}{2}\le \arg (\la_k-\la_j)\le -\theta_{j,j+1}+\frac{3\pi}{2}\mod 2\pi.
\end{equation}
By adding this inequality and (\ref{es7}), we obtain that
\begin{equation}\label{es11}
\frac{\pi}{2}< \arg (\la_k-\la_j)z<\frac{3\pi}{2}\mod 2\pi,
\end{equation}
which implies (\ref{es8}). Proposition \ref{T_j} is proved.
\end{proof}

For the future use, observe that if $k\not=j-1,\,j,\,j+1,$ then
inequality (\ref{es9}) is strict and hence there exists $\ep>0$
such that
\begin{equation}\label{es12}
\frac{\pi}{2}+\ep< \arg (\la_k-\la_j)z<\frac{3\pi}{2}-\ep.
\end{equation}
This gives that for some $c>0$,
\begin{equation}\label{es13}
e^{(\la_k-\la_j)z}=O(e^{-c|z|}),\quad |z|\to\infty \qquad k\not=j-1,\,j,\,j+1,
\end{equation}
uniformly in the closed sector $\overline{\mathcal U_j}$. From (\ref{es3})
we obtain that for some $\ep>0$,
\begin{equation}\label{es14}
\frac{\pi}{2}< \arg \,(\la_{j+1}-\la_j)z<\frac{3\pi}{2}-\ep,\quad z\in \mathcal U_j,
\end{equation}
and from (\ref{es5}), that
\begin{equation}\label{es15}
 \arg \,(\la_{j+1}-\la_j)z=\frac{\pi}{2},\quad z\in \mathcal S_{j,j+1}.
\end{equation}
This gives that for some $c>0$,
\begin{equation}\label{es16}
e^{(\la_{j+1}-\la_j)z}=O(e^{-cd_{j,j+1}(z)}),\quad d_{j,j+1}(z)
\equiv\dist(z,\mathcal S_{j,j+1})\to\infty;\qquad z\in \mathcal U_j.
\end{equation}
Similarly,
\begin{equation}\label{es17}
e^{(\la_{j-1}-\la_j)z}=O(e^{-cd_{j-1,j}(z)}),\quad 
d_{j-1,j}(z)\to\infty;\qquad z\in \mathcal U_j.
\end{equation}
Estimates (\ref{es13}), (\ref{es16}), and (\ref{es17}) imply that
there exist large numbers $r_0,R_0>0$ such that for $j=1,\ldots,m$,
\begin{equation}\label{es18}
\sum_{k:\; k\not=j}\left|
\frac{c_k}{c_j}\,e^{(\la_k-\la_j)z}\right|<\frac{1}{2}\,,
\qquad z\in \mathcal U_j(r_0,R_0),
\end{equation}
where
\begin{equation}\label{es19}
\mathcal U_j(r_0,R_0)=\{z\in \mathcal U_j:\; |z|>R_0,\;\dist(z,\mathcal S_{j,j+1})> r_0,\; 
\dist(z,\mathcal S_{j-1,j})> r_0\}.
\end{equation}
We will call $\mathcal U_j(r_0,R_0)$ the {\it $j$-th one-term domination
domain}. When $z\in \mathcal U_j(r_0,R_0)$,
the term $c_je^{\la_jz}$ dominates in $f(z)$ the other terms.
Define
\begin{equation}\label{es20}
\mathcal U(r_0,R_0)=\bigcup_{j=1}^m \mathcal U_j(r_0,R_0).
\end{equation}
Define also
\begin{equation}\label{es21}
\mathcal S_{j,j+1}(r_0,R_0)=\{z:\; |z|>R_0,\;\dist(z,\mathcal S_{j,j+1})\le r_0 \},
\qquad j=1,\ldots,m.
\end{equation}
We will assume that $R_0$ is big enough so that 
\begin{equation}\label{es22}
\mathcal S_{j,j+1}(r_0,R_0)\bigcap \mathcal S_{j-1,j}(r_0,R_0)=\emptyset,
\qquad j=1,\ldots,m.
\end{equation}
We will call $\mathcal S_{j,j+1}(r_0,R_0)$ the {\it $(j,j+1)$-st two-term domination
strip}. Define
\begin{equation}\label{es23}
\mathcal S(r_0,R_0)=\bigcup_{j=1}^m \mathcal S_{j,j+1}(r_0,R_0).
\end{equation}

\begin{prop} \label{one-term}
{\rm (Absence of zeros of $f$ in the one-term domination domains).}
There exists $r_0,\;R_0>0$ such that
\begin{equation}\label{dom9}
\sum_{k=1}^M c_ke^{\la_k z}\not=0,
\qquad z\in \mathcal U(r_0,R_0).
\end{equation}
\end{prop}

\begin{proof}
The proof follows from (\ref{es6}) and (\ref{es18}).
\end{proof}

Proposition \ref{one-term} implies that
 all the large zeros of $f$ are concentrated in the  two-term domination
strips, $\mathcal S_{j,j+1}(r_0,R_0)$.
To describe these zeros consider the two-term equation,
\begin{equation}\label{dom12}
f_0(z)\equiv c_je^{\la_j z}+c_{j+1}e^{\la_{j+1} z}=0.
\end{equation}
By the linear change  of variable,
\begin{equation}\label{dom12a}
u=\frac{(\la_{j+1}-\la_j)z}{2i}+\frac{1}{2i}\log\frac{c_{j+1}}{c_j},
\end{equation}
we reduce $f_0$ to
\begin{equation}\label{dom12b}
f_0(z)=2\sqrt{c_jc_{j+1}}e^{\frac{(\la_{j+1}+\la_j)z}{2}}\cos u.
\end{equation}
Therefore, the general solution to equation (\ref{dom12}) is $u=\frac{\pi}{2}+\pi l$, or
\begin{equation}\label{dom13}
z=z^0(j,j+1;l)\equiv \al_{j,j+1}+l\tau_{j,j+1},\qquad n\in\Z,
\end{equation}
where
\begin{equation}\label{dom14}
\al_{j,j+1}=\frac{\pi i- \log\frac{c_{j+1}}{c_j}}{\la_{j+1}-\la_j}\,,
\qquad \tau_{j,j+1}=\frac{2\pi i}{\la_{j+1}-\la_j}\,.
\end{equation}
Observe that
\begin{equation}\label{dom15}
\arg \tau_{j,j+1}=\theta_{j,j+1}\,.
\end{equation}
Now we can describe the zeros of $f$ in the two-term domination
strips. We will use the following general proposition. Denote
\begin{equation}\label{dom16}
D(z_0,r)=\{z:\;|z-z_0|<r\}, \qquad r>0.
\end{equation}

\begin{prop}\label{hurwitz} Let $f(z)=f_0(z)+f_1(z)$ where
$f_0,\;f_1$ are analytic functions in the disk $D(z_0,r)$, $r>0$. Suppose that
\begin{itemize}
\item
$f_0(z_0)=0$,
\item
$|f_0(z)|\ge A|z-z_0|$, $\forall\,z\in D(z_0,r)$, where $A>0$,
\item
$|f_1(z)|\le \ep$, $z\in D(z_0,r)$, $\ep>0$.
\end{itemize}
Then if $r_0\equiv \frac{2\ep}{A}< r$, then 
there is a unique simple zero of $f$ in the disk $D(z_0,r_0)$.
\end{prop}

\begin{proof}
 For $|z-z_0|=r_0$, $|f_0(z)|\ge 2\ep>|f_1(z)|$,
hence $f$ has a unique simple zero in $D(z_0,r_0)$ by the
Rouch\'e theorem. Proposition \ref{hurwitz} is proved.
\end{proof}

With the help of Proposition \ref{hurwitz} we prove the following result.

\begin{prop} \label {two-term}
{\rm (Zeros of $f$ in the two-term domination strips).} 
There exist $r_0,\; R_0>0$ such that
all zeros $z_k$ of exponential sum (\ref{es1}) in $\mathcal S_{j,j+1}(r_0,R_0)$
are simple and close to zeros (\ref{dom13}),
so that for some $l=l(k)>0$, 
\begin{equation}\label{dom17}
|z_k-z^0(j,j+1;l)|=O(e^{-cl}),\quad c>0, 
\end{equation}
and for each $z^0(j,j+1;l)\in \mathcal S_{j,j+1}(r_0,R_0)$, there is a zero
$z_k$ of $f$ satisfying (\ref{dom17}).
\end{prop}

\begin{proof}
 From  (\ref{es13}) and (\ref{es17}) we obtain that
if $z\in \mathcal S_{j,j+1}(r_0,R_0)$, then for some $c>0$,
\begin{equation}\label{dom18}
e^{(\la_k-\la_j)z}=O(e^{-c|z|}),\quad
e^{(\la_k-\la_{j+1})z}=O(e^{-c|z|}),\quad |z|\to\infty; \qquad k\not=j,\,j+1.
\end{equation}
Let us write equation $f(z)=0$ as 
\begin{equation}\label{dom19}
f_0(z)+f_1(z)=0,\qquad f_1(z)=\sum_{k\not=j,\,j+1}c_ke^{\la_k z}.
\end{equation}
Then (\ref{dom18}) implies that if $z\in \mathcal S_{j,j+1}(r_0,R_0)$, then
\begin{equation}\label{dom20}
e^{-\frac{(\la_{j+1}+\la_j)z}{2}}\,f_1(z)=O(e^{-c|z|}),\quad c>0,\qquad |z|\to\infty, 
\end{equation}
and under transformation (\ref{dom12a}) equation (\ref{dom19}) becomes
\begin{equation}\label{dom21}
\cos u +g_1(u)=0, \qquad g_1(u)=O(e^{-c_0\Re u}),\quad c_0>0;\qquad 
\Re u\to\infty.
\end{equation}
Proposition \ref{hurwitz} implies that 
for any $a>0$ there exists $b>0$ such that 
all zeros of the latter equation in the region
\begin{equation}\label{dom22}
\{u:\; |\Im u|<a,\;\Re u>b\}
\end{equation}
are simple and of the form,
\begin{equation}\label{dom23}
u^0(l)=\frac{\pi}{2}+\pi l+O(e^{-c_0\Re u}).
\end{equation}
This implies (\ref{dom17}). Proposition \ref{two-term} is proved.
\end{proof}

We will call $z_k\in \mathcal S(r_0,R_0)$, the {\it zeros of the main series}.
We summarize the results of this section as follows.

\begin{theo} \label{zeros_exp_sum}
{\rm (Zeros of the exponential sum.)} 
Suppose that the numbers $\la_j$ satisfy Condition P.
Then there exists $r_0,\;R_0>0$ 
such that all the zeros of $f$ belong to one of the following categories:
\begin{itemize}
\item
$|z_k|\le R_0$ (finite zeros)
\item
$z_k\in \mathcal S(r_0,R_0)$, described by formula (\ref{dom17}) 
(zeros of the main series).
\end{itemize}   
\end{theo}

\section{Zeros of sections of exponential sums}\label{ZSES}

Denote by $f_n(z)$ the section of the exponential sum $f(z)$,
\begin{equation}\label{sec1}
f_n(z)=\sum_{k=0}^n \frac {f^{(k)}(0)z^k}{k!}\,.
\end{equation}
By (\ref{es1}),
\begin{equation}\label{sec2}
f_n(z)=\sum_{k=0}^n a_kz^k,
\qquad a_k=\sum_{j=1}^M \frac{c_j\la_j^k}{k!}\,.
\end{equation}
Our main goal will be to decribe the zeros of the polynomial $f_n(z)$,
\begin{equation}\label{sec3}
f_n(z)=0,
\end{equation}
as $n\to\infty$. We expect that
as $n\to\infty$ some of the zeros of $f_n(z)$
approach the zeros of $f(z)$. We call them the {\it genuine zeros}
of $f_n$. We divide the genuine zeros into 
{\it finite zeros} and {\it zeros
of the main series}, in accordance with Theorem \ref{zeros_exp_sum}.
 But there is also a family of other zeros, which go to infinity
as $n\to\infty$.
We call them the {\it spurious zeros}. In addition, there will
be a relatively small number of intermediate zeros. We call
them the {\it transitional zeros}. In the following sections, we will describe
all these zeros of $f_n$.

\section {Finite zeros} 

It will be more convenient for us
to consider, instead of (\ref{sec3}), the equation
\begin{equation}\label{ms1}
f_{n-1}(z)=0.
\end{equation}
We rewrite it as
\begin{equation}\label{ms2}
f(z)=\sum_{k=n}^\infty a_kz^k=a_nz^n\sum_{k=0}^\infty\frac{a_{n+k}}{a_n}z^k.
\end{equation}
In addition to Condition P, we will assume the following condition:

\vskip 2mm

\noindent
{\bf Condition Q.}
{\it One of $|\la_j|$'s,
say $|\la_1|$, is bigger than the others. }

\vskip 2mm

By the change of variables, $\la_1 z\to z$, we
can reduce $\la_1$ to 1, so we will assume that
\begin{equation}\label{ms3}
\la_1=1>|\la_j|,\qquad j=2,\ldots,N.
\end{equation}
Also we can assume that
\begin{equation}\label{ms3a}
c_1=1.
\end{equation}
In this case, by (\ref{sec2}), as $n\to\infty$,
\begin{equation}\label{ms4}
a_n=\frac{1}{n!}(1+O(q^n)),\qquad 0<q<1.
\end{equation}
Therefore, equation (\ref{ms2}) reads
\begin{equation}\label{ms5}
f(z)=\frac{z^n}{n!}(1+O(q^n))\left[1+\sum_{k=1}^\infty
\frac{z^k(1+O(q^{n+k}))}{(n+1)\ldots(n+k)(1+O(q^n))}\right].
\end{equation}
By the Stirling formula,
\begin{equation}\label{ms6}
n!=\frac{n^n}{e^n}\sqrt{2\pi n}\,e^{\frac{\theta}{12n}},\qquad 0<\theta<1,
\end{equation}
hence we can rewrite (\ref{ms5}) as
\begin{equation}\label{ms7}
f(z)=\frac{e^n z^n e^{-\frac{\theta}{12n}}}{n^n\sqrt{2\pi n}}(1+O(q^n))
\left[1+\sum_{k=1}^\infty
\frac{z^k(1+O(q^{n+k}))}{(n+1)\ldots(n+k)(1+O(q^n))}\right],
\end{equation}
where the $O$-terms are independent of $z$.
If $z$ is bounded, $|z|<R_0$, then the right hand side is $O(e^{-An})$
as $n\to\infty$ for 
any $A>0$. Hence the zeros, with multiplicities, of $f_n$ are close to those of $f$.
More precisely, the following proposition holds.

\begin{prop} \label{finite_zeros} {\rm (Finite zeros of $f_{n-1}$).} Let $R_0>0$ be
a fixed number such that $f$ has no zeros on the circle $|z|=R_0$. Then for large $n$,
there is a one-to-one correspondence between zeros $z_k\in D(0,R_0)$ of $f$,
and zeros
$z_k(n)\in D(0,R_0)$ of $f_{n-1}$ such that
\begin{equation}\label{ms8}
z_k(n)-z_k=O(e^{-An}),\qquad n\to\infty,
\end{equation}
for any $A>0$. Here any zero of multiplicity $p$ is counted as $p$ zeros.
\end{prop}

\section{Zeros of the main series}\label{ZMS}

Consider now zeros of $f$ 
in the two-term domination strip $\mathcal S_{j,j+1}(r_0,R_0)$.
Let us write $f$ as
\begin{equation}\label{ms9}
f(z)=f_0(z)+f_1(z),
\qquad f_0(z)=c_je^{\la_j z}+c_{j+1}e^{\la_{j+1} z},
\end{equation}
so that $f_0$ dominates $f_1$ in $S_{j,j+1}(r_0,R_0)$. With the help of
substitution (\ref{dom12a}),  
we reduce $f_0$ to form (\ref{dom12b}). In (\ref{dom12a}), (\ref{dom12b})
we choose the branch for $\log\frac{c_{j+1}}{c_j}$ and $\sqrt{c_jc_{j+1}}$ as follows: 
if $c_j=r_je^{i\theta_j}$, $-\pi<\theta_j\le \pi$, $j=1,\ldots,m$,
then we define
\begin{equation}\label{ms11a}
\log\frac{c_{j+1}}{c_j}=\ln \frac{r_{j+1}}{r_j}+i(\theta_{j+1}-\theta_j),
\qquad
\sqrt{c_jc_{j+1}}=\sqrt{r_jr_{j+1}}e^{i\frac{\theta_j+\theta_{j+1}}{2}}
\end{equation}
Under (\ref{dom12a}), equation (\ref{ms7})  reduces to the form,
\begin{equation}\label{m12}
\begin{aligned}
\cos u+O(e^{-c\Re u})&=\frac{e^nz^ne^{-\frac{(\la_{j+1}+\la_j)z}{2}}
e^{-\frac{\theta}{12n}}}
{2\sqrt{c_jc_{j+1}}\,n^n\sqrt{2\pi n}}(1+O(q^n))\\
&\times\left[1+\sum_{k=1}^\infty
\frac{z^k(1+O(q^{n+k}))}{(n+1)\ldots(n+k)(1+O(q^n))}\right].
\end{aligned}
\end{equation}
After the rescaling,
\begin{equation}\label{ms13}
z=n\z,
\end{equation}
we obtain the equation,
\begin{equation}\label{m14}
\begin{aligned}
\cos u+O(e^{-c\Re u})&=\frac{e^n\z^ne^{-\frac{(\la_{j+1}+\la_j)n\z}{2}}
e^{-\frac{\theta}{12n}}}
{2\sqrt{c_jc_{j+1}}\,\sqrt{2\pi n}}(1+O(q^n))\\
&\times\left[1+\sum_{k=1}^\infty
\frac{n^k\z^k(1+O(q^{n+k}))}{(n+1)\ldots(n+k)(1+O(q^n))}\right].
\end{aligned}
\end{equation}
Let us discuss the condition when the right hand side in this
equation is $o(1)$ as $n\to\infty$.
As a first approximation to this, consider 
the {\it critical radius} $r_c=r_c(j,j+1)>0$ on the ray $\{\z:\;\arg\z=\theta_{j,j+1}\},$
as a solution of the equation
\begin{equation}\label{ms15}
e\left|\z e^{-\frac{(\la_{j+1}+\la_j)\z}{2}}\right|=1,\qquad 
\z=r_ce^{i\theta_{j,j+1}},
\end{equation}
on the interval $0<r_c<1$.
This equation can be rewritten as
\begin{equation}\label{ms16}
r_c e^{1+r_c x_{j,j+1}}=1,
\end{equation}
where
\begin{equation}\label{ms17}
x_{j,j+1}=\frac{|\la_{j+1}+\la_j|}{2}\cos \beta_{j,j+1},
\end{equation}
and
\begin{equation}\label{ms18}
\beta_{j,j+1}=\arg(\la_{j+1}+\la_j)+\theta_{j,j+1}-\pi
=-\frac{\pi}{2}+\arg \frac{\la_{j+1}+\la_j}{\la_{j+1}-\la_j}.
\end{equation}

\begin{prop}\label{critical_radius} {\rm (Existence of the critical radius).}
There exists a unique solution of equation (\ref{ms15}) on the interval $0<r_c<1$.
\end{prop}

\begin{proof}
Observe that (\ref{ms3}) implies that
\begin{equation}\label{ms19}
-1<x_{j,j+1}<1.
\end{equation}
From this condition we obtain that the function
\begin{equation}\label{ms20}
g(r)=r e^{1+r x_{j,j+1}}
\end{equation}
is increasing on $[0,1]$. Indeed,
\begin{equation}\label{ms21}
g'(r)=(1+rx_{j,j+1}) e^{1+r x_{j,j+1}}>0,\qquad 0\le r\le 1.
\end{equation}
Also, $g(0)=0$ and $g(1)>1$,
hence equation (\ref{ms16}) has a unique solution on the
interval $0<r_c<1$, QED.
\end{proof}

Let $0<r^*<1$ be a solution of the equation,
\begin{equation}\label{ms21a}
r^* e^{1+r^*}=1. 
\end{equation}
We have that
\begin{equation}\label{ms21b}
r^*=0.27846\ldots 
\end{equation}
From (\ref{ms19}) we obtain that $g(r^*)<1$, hence
\begin{equation}\label{ms21c}
r^*<r_c<1.
\end{equation}

In the disk $|\z|\le r_c<1$, the function
\begin{equation}\label{ms21d}
1+\sum_{k=1}^\infty
\frac{n^k\z^k(1+O(q^{n+k}))}{(n+1)\ldots(n+k)(1+O(q^n))}
\end{equation}
is well approximated by
\begin{equation}\label{ms21e}
1+\sum_{k=1}^\infty \z^k=\frac{1}{1-\z},
\end{equation}
so that
\begin{equation}\label{ms22}
\left|1+\sum_{k=1}^\infty
\frac{n^k\z^k(1+O(q^{n+k}))}{(n+1)\ldots(n+k)(1+O(q^n))}-\frac{1}{1-\z}\right|
=O(n^{-1}).
\end{equation}
Therefore, for $z\in \mathcal S_{j,j+1}\cap\{|z|\le nr_c\}$,
equation (\ref{m14}) reduces to
\begin{equation}\label{ms23}
\cos u+O(e^{-c\Re u})=\frac{e^n\z^ne^{-\frac{(\la_{j+1}+\la_j)n\z}{2}}}
{2\sqrt{c_jc_{j+1}}\,\sqrt{2\pi n}\,(1-\z)}(1+O(n^{-1})).
\end{equation}
Introduce the $n$-th critical radius, $r_c^n=r_c^n(j,j+1)>0$, 
on the ray $\{\z:\;\arg\z=\theta_{j,j+1}\},$
as a solution of the equation 
\begin{equation}\label{ms24}
\left|\frac{e\z e^{-\frac{(\la_{j+1}+\la_j)\z}{2}}}
{\left[2\sqrt{c_jc_{j+1}}\,\sqrt{2\pi n}\,(1-\z)\right]^{\frac{1}{n}}}\right|=1\,,\qquad 
\z=\z_c^n\equiv r_c^n e^{i\theta_{j,j+1}},
\end{equation}
on the interval $0<r_c^n<1$. Observe that $r_c^n$ is a small
correction to $r_c$,
\begin{equation}\label{ms25}
r_c^n=r_c+O(n^{-1}\ln n).
\end{equation}

\begin{theo} \label{main_series}
{\rm (Zeros of the main series).} 
There exists a (big) number  $R_1>0$ such that 
for any zero $z_k$ of $f$ in the region,
\begin{equation}\label{ms26}
\mathcal S_{j,j+1}(r_0,R_0,R_1;n)=\mathcal S_{j,j+1}(r_0,R_0)\bigcap\{ z:\;
|z|<n r_c^n-R_1\},\quad r_c^n=r_c^n(j,j+1),
\end{equation}
there exists a unique zero
$z_k(n)$ of $f_{n-1}$ such that
\begin{equation}\label{ms27}
z_k(n)-z_k=O(e^{-\ga (n r_c^n-|z_k|)}),\qquad n\to\infty,
\end{equation}
where $\ga>0$ is independent of $n$. There exists $N>0$ such that $\forall\,n>N$,
the zeros $z_k(n)$, described by (\ref{ms27}), exhaust all the zeros
of $f_{n-1}$ in the region $\mathcal S_{j,j+1}(r_0,R_0,R_1;n)$.
\end{theo}

\begin{proof}
 Define
\begin{equation}\label{ms28}
\be_n(\z)=\frac{e\z e^{-\frac{(\la_{j+1}+\la_j)\z}{2}}}
{\left[2\sqrt{c_jc_{j+1}}\,\sqrt{2\pi n}\,(1-\z)\right]^{\frac{1}{n}}},
\qquad |\z|<r^n_c<1.
\end{equation}
Then 
\begin{equation}\label{ms29}
\lim_{n\to\infty}\be_n(\z)=\be(\z)\equiv
e\z e^{-\frac{(\la_{j+1}+\la_j)\z}{2}},
\qquad |\z|<r^n_c.
\end{equation}
and by (\ref{ms24}),
\begin{equation}\label{ms30}
|\be_n(r_c^ne^{i\theta_{j,j+1}})|=1.
\end{equation}
The function $g(r)=|\be(r e^{i\theta_{j,j+1}})|$ is strictly increasing 
on the interval $0<r<1$ and this, together with (\ref{ms29}), (\ref{ms30}),
 implies that for large $n$,
\begin{equation}\label{ms30a}
|\be_n(r e^{i\theta_{j,j+1}})|\le e^{c(r-r_c^n)},\quad c>0;
\qquad 0<r\le r_c^n.
\end{equation}

Equation (\ref{ms23}) is of the form
\begin{equation}\label{ms31}
\al(u)=\ga_n(z),
\end{equation}
where $u$ and $z$ are related as in (\ref{dom12a}),
$\al(u)$ an entire function such that in the
region $\{u:\; |\Im u|<a,\; \Re u>b\}$, $a,b>0$,
\begin{equation}\label{ms32}
\al(u)=\cos u+O(e^{-c\Re u}),
\end{equation}
and $\ga_n(z)$ is an entire function such that
\begin{equation}\label{ms33}
\ga_n(z)=\be_n(\z)^n(1+O(n^{-1})),\qquad \z=\frac{z}{n}\,.
\end{equation}
From (\ref{ms30a}) and (\ref{ms29}) we obtain that in the
region $\{u:\; |\Im u|<a,\; \Re u>b\}$,
\begin{equation}\label{ms34}
|\ga_n(z)|<C|\be_n(\z)|^n<C e^{c(|z|-nr_c^n)},\quad C,c>0; \qquad |z|< nr_c^n.
\end{equation}
By Proposition \ref{hurwitz}  this implies that for a given $a>0$
there exists $b>0$ and $R_0>0$ such that all the zeros, $u_k(n)$, 
of equation (\ref{ms31}) in the
region $\{u:\; |\Im u|<a,\; \Re u>b,\; |z|<nr_c^n-R_1\}$,
are simple and close to the zeros, $u_k$, of the function $\al(u)$, so that
\begin{equation}\label{ms35}
u_k(n)-u_k=O(e^{-c (n r_c^n-|z_k|)}),\qquad n\to\infty,
\end{equation}
which implies (\ref{ms27}). Theorem \ref{main_series} is proved.
\end{proof}

We will call the zeros $z_k(n)$ satisfying (\ref{ms27}), the {\it zeros 
of the  main series} of $f_{n-1}$. As $n\to\infty$, they approach
the zeros $z_k$ of $f$. This is true also for the finite zeros of
Proposition \ref{finite_zeros}. 

\section{Spurious zeros}\label{SPZ}

We will construct a sequence of spurious zeros of $f_{n-1}(z)$ in the 
$j$-th one-term domination sector, $\mathcal U_j(r_0,R_0)$. Let us first discuss the
construction informally. 

{\it Construction of the rosette.}
In $\mathcal U_j(r_0,R_0)$,
\begin{equation}\label{sp1}
f(z)=c_je^{\la_j z}(1+O(e^{-d_j(z)})),
\end{equation}
where
\begin{equation}\label{sp1a}
d_j(z) =c\min\{\dist(z,S_{j,j+1}),\;\dist(z,S_{j-1,j}),\;|z|\},
\quad c>0,
\end{equation}
hence equation (\ref{ms7}), which is equivalent to the equation $f_{n-1}(z)=0$, reads
\begin{equation}\label{sp2}
1+O(e^{-d_j(z)})=\frac{e^n z^n e^{-\la_j z}
e^{-\frac{\theta}{12n}}(1+O(q^n))}{n^nc_j\sqrt{2\pi n}}
\left[1+\sum_{k=1}^\infty
\frac{z^k(1+O(q^{n+k}))}{(n+1)\ldots(n+k)(1+O(q^n))}\right],
\end{equation}
or after the scaling $z=n\z$,
\begin{equation}\label{sp3}
\begin{aligned}
1+O(e^{-d_j(z)})&=\frac{e^n \z^n e^{-\la_j n\z}
e^{-\frac{\theta}{12n}}(1+O(q^n))}{c_j\sqrt{2\pi n}}\\
&\times\left[1+\sum_{k=1}^\infty
\frac{n^k\z^k(1+O(q^{n+k}))}{(n+1)\ldots(n+k)(1+O(q^n))}\right].
\end{aligned}
\end{equation}
By taking the $n$-th root, we obtain the equation,
\begin{equation}\label{sp4}
\begin{aligned}
\om_q(1+O(e^{-d_j(z)}))^{\frac{1}{n}}&=\frac{e \z e^{-\la_j \z}
e^{-\frac{\theta}{12n^2}}(1+O(q^n))^{\frac{1}{n}}}{(c_j\sqrt{2\pi n})^{\frac{1}{n}}}\\
&\times\left[1+\sum_{k=1}^\infty
\frac{n^k\z^k(1+O(q^{n+k}))}{(n+1)\ldots(n+k)(1+O(q^n))}\right]^{\frac{1}{n}},
\end{aligned}
\end{equation}
where $\om_q=e^{\frac{2\pi q i}{n}}$, $q=0,1,\ldots,n-1$. As an approximation to this equation,
consider the equation,
\begin{equation}\label{sp5}
e \z e^{-\la_j \z}=\om_q.
\end{equation}
By taking the absolute value of the both sides, we obtain the equation
of the curve on the complex plane,
\begin{equation}\label{sp6}
e \left|\z e^{-\la_j \z}\right|=1.
\end{equation}
For $\la_1=1$ it reduces to the Szeg\"o equation,
\begin{equation}\label{sp7}
e \left|\z e^{-\z}\right|=1,
\end{equation}
and for $\la_j=0$, to the equation of the circle,
\begin{equation}\label{sp8}
|\z|=e^{-1}.
\end{equation}
If $0<\la_j<1$, it can be reduced to the equation,
\begin{equation}\label{sp9}
e \left|\xi e^{-|\la_j| \xi}\right|=1,\qquad \xi=e^{i\arg \la_j}\z\,.
\end{equation}
If $0<c<1$, the set of solutions of the equation
\begin{equation}\label{sp10}
e \left|\xi e^{-c \xi}\right|=1
\end{equation}
on the complex plane consists of two  analytic
curves: $\Ga=\Ga(c)$, inside of the
unit circle, and $\Ga_0=\Ga_0(c)$, outside of the unit circle. 
In the polar coordinates, $\xi=r e^{i\theta}$, equation (\ref{sp10})
reads
\begin{equation}\label{sp11}
\cos\theta=g(r),\qquad g(r)=\frac{1+\ln r}{c r}\,.
\end{equation}
Observe that
\begin{equation}\label{sp12}
g'(r)=-\frac{\ln r}{c r^2},
\end{equation}
and $g(r)$ attains a maximum at $r=1$, with $g(1)=\frac{1}{c}>1$.
Also, $g(0)=-\infty$ and $g(r)$ is increasing on $(0,1]$.
Hence equation (\ref{sp11}) has a unique solution in the
interval $0<r<1$ for any $\theta$, and this solution determines
the oval $\Ga(c)$. We will call (\ref{sp10}) the {\it generalized 
Szeg\"o equation} and $\Ga(c)$ the {\it generalized Szeg\"o curve.}
Figure 4 depicts the generalized Szeg\"o curve for $c=0.9$.
In the Cartesian coordinates the equation of $\Ga(c)$ has the form,
\begin{equation}\label{sp12a}
y=\pm \sqrt{e^{2cx-2}-x^2}\,.
\end{equation}

\begin{center}
\begin{figure}[h]\label{figure4}
\begin{center}
\scalebox{0.5}{\includegraphics{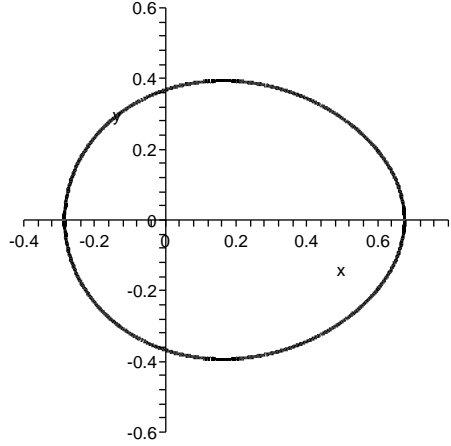}}
\end{center}
   \caption{The generalized Szeg\"o curve for $c=0.9$.}
   \end{figure}
\end{center}

The function
\begin{equation}\label{sp13}
h(\xi)=e\xi e^{-c\xi},
\end{equation}
is entire and it conformally maps the interior of the curve $\Ga(c)$
onto the unit disk,  
\begin{equation}\label{sp14}
h:\;{\rm Int}\,\Ga(c)\to D(0,1)=\{ z:\;|z|< 1\}.
\end{equation}
The preimage, with respect to $h$, of the uniform probability measure on the unit circle,
$(2\pi)^{-1}d\theta$, is the measure $d\mu_{{\rm max}}(\theta)$ of the
maximal entropy on $\Ga(c)$.

We will denote the curve
\begin{equation}\label{sp14a}
\mathcal G(\la_j)=\{\z:\; e \left|\z e^{-\la_j \z}\right|=1,\;|\z|<1\},
\end{equation}
so that
\begin{equation}\label{sp14b}
\mathcal G(\la_j)=e^{-i\arg \la_j}\Ga(|\la_j|).
\end{equation}

Recall that the sector $\mathcal U_j(r_0,R_0)$ is given by the inequalities,
\begin{equation}\label{sp15}
\theta_{j,j+1}<\arg \z 
<\theta_{j-1,j}\mod 2\pi,
\qquad \theta_{j,j+1}=-\arg(\la_{j+1}-\la_j)+\frac{\pi}{2}.
\end{equation}
According to (\ref{sp9}), this implies that if $\la_j\not=0$ then
\begin{equation}\label{sp16}
\begin{aligned}
\al_{j,j+1}&<\arg \xi 
<\al_{j-1,j} \mod 2\pi;\\
\al_{j,j+1}&=-\arg(\la_{j+1}-\la_j)+\frac{\pi}{2}+\arg\la_j.
\end{aligned}
\end{equation}
This condition holds also for $\la_j=0$, if we take the agreement
that $\arg \la_j=0$ for $\la_j=0$. Consider now the arc,
\begin{equation}\label{sp17}
\Ga_j=\{\xi:\; \xi\in\Ga(|\la_j|),\;
\al_{j,j+1}<\arg \xi 
<\al_{j-1,j} \mod 2\pi\},
\end{equation}
on $\Ga(|\la_j|)$, and the arc,
\begin{equation}\label{sp18}
\begin{aligned}
\mathcal G_j=e^{-i\arg\la_j}\Ga_j&=\{\z:\; \z \in e^{-i\arg\la_j}\Ga(|\la_j|),\;
\theta_{j,j+1}<\arg \z 
<\theta_{j-1,j}\mod 2\pi\},
\end{aligned}
\end{equation}
on $\mathcal G(\la_j)$.
The arc $\mathcal G_j\subset \mathcal G(\la_j)$ goes 
from one side of the sector $\mathcal U_j(r_0,R_0)$ to another.
More precisely, we have the following statement. Consider the points
\begin{equation}\label{sp19}
\z_c(j,j+1)=r_c(j,j+1)e^{i\theta_{j,j+1}},\qquad j=1,\ldots,m.
\end{equation}

\begin{prop} \label{Sigma_j}
The arc $\mathcal G_j$ connects the point $\z_c(j-1,j)$ to the point $\z_c(j,j+1)$.
\end{prop}

\begin{proof} 
From (\ref{dom12a}) we have that if $\arg z=\theta_{j,j+1}$ then
\begin{equation}\label{sp20}
\la_j z=\frac{(\la_{j+1}+\la_j)z}{2}-iu,\qquad u\in\R,
\end{equation}
hence
\begin{equation}\label{sp21}
|e^{-\la_j \z}|=\left| e^{-\frac{(\la_{j+1}+\la_j)\z}{2}}\right|,\qquad \z=\frac{z}{n}\,,
\end{equation}
and equations (\ref{ms15}) and (\ref{sp6}) coincide. This proves that 
$\z_c(j,j+1)\in \mathcal G_j$. The relation $\z_c(j-1,j)\in \mathcal G_j$ is
established in the same way. Proposition \ref{Sigma_j} is proved.
\end{proof}

The {\it rosette} $\mathcal H$ is, by definition, the union of the arcs $\mathcal G_j$
 and the finite rays,
\begin{equation}\label{sp22}
\mathcal R_{j,j+1}=\{ z=r e^{i\theta}:\; \theta=\theta_{j,j+1},\; 0\le r\le r_c(j,j+1)\},
\end{equation}
$j=1,\ldots,m$, so that
\begin{equation}\label{sp22a}
\mathcal H=\mathcal R\bigcup \mathcal G,
\end{equation}
where
\begin{equation}\label{sp22b}
\mathcal R=\bigcup_{j=1}^m\mathcal R_{j,j+1}
\end{equation}
and
\begin{equation}\label{sp22c}
\mathcal G=\bigcup_{j=1}^m\mathcal G_j.
\end{equation}
By definition,
the $j$-th {\it petal}, $\mathcal P_j$, of the rosette $\mathcal H$ is the region bounded 
by the rays $\mathcal R_{j,j+1}$, $\mathcal R_{j-1,j}$, and the arc $\mathcal G_j$.

{\it Construction of the $n$-th rosette.} As a better, than (\ref{sp5}),
approximation to equation (\ref{sp4}), consider the equation,
\begin{equation}\label{sp23}
\frac{e \z e^{-\la_j \z}}{\left[c_j\sqrt{2\pi n}\,(1-\z)\right]^{\frac{1}{n}}}=\om_q.
\end{equation}
By taking the absolute value of the both sides, we obtain the equation,
\begin{equation}\label{sp23a}
\frac{e \left|\z e^{-\la_j \z}\right|}{\left[c_j\sqrt{2\pi n}\,|1-\z|\right]^{\frac{1}{n}}}=1,
\end{equation}
Consider the curve,
\begin{equation}\label{sp23b}
\mathcal G^n(\la_j)=
\left\{\z:\; \frac{e \left|\z e^{-\la_j \z}\right|}{\left[c_j\sqrt{2\pi n}\,|1-\z|\right]^{\frac{1}{n}}}=1,
\;|\z|<1\right\}
\end{equation}
and the arc $\mathcal G_j^n\subset \mathcal G^n(\la_j)$, which goes from the point
\begin{equation}\label{sp24}
\z_c^n(j,j+1)=r_c^n(j,j+1)e^{i\theta_{j,j+1}}
\end{equation}
to the point $\z_c^n(j-1,j)$. Observe that if $|\la_j|<1$ then 
for large $n$, the curve $\mathcal G^n(\la_j)$ lies {\it outside} of the
curve $\mathcal G(\la_j)$, and
\begin{equation}\label{sp24a}
\dist(\mathcal G^n(\la_j),\mathcal G(\la_j))=\frac{c\ln n}{n}(1+o(1)),\quad
n\to\infty;\qquad c>0.
\end{equation}
Also, if $\de>0$ is fixed, then for large $n$,
the curve $\mathcal G^n(1)\setminus D(1,\de)$ lies outside of the
curve $\mathcal G(1)\setminus D(1,\de)$, so that the equations 
of $\mathcal G^n(1)\setminus D(1,\de)$ and $\mathcal G(1)\setminus D(1,\de)$ in the polar
coordinates, $\rho=\ga^n(\theta)$ and $\rho=\ga(\theta)$, satisfy $\ga^n(\theta)>\ga(\theta)$, and
\begin{equation}\label{sp24b}
\dist(\mathcal G^n(1)\setminus D(1,\de),\mathcal G(1)\setminus D(1,\de))=\frac{c(\de)\ln n}{n}(1+o(1)),\quad
n\to\infty;\qquad c(\de)>0.
\end{equation}

The {\it $n$-th rosette}, $\mathcal H^n$, is, by definition,
the union of the arcs $\mathcal G_j^n$
 and the finite rays,
\begin{equation}\label{sp25}
\mathcal R_{j,j+1}^n=\{ z=r e^{i\theta}:\; \theta=\theta_{j,j+1},\; 0\le r\le r_c^n(j,j+1)\},
\end{equation}
$j=1,\ldots,m$, so that
\begin{equation}\label{sp25a}
\mathcal H^n=\mathcal R^n\bigcup \mathcal G^n,
\end{equation}
where
\begin{equation}\label{sp25b}
\mathcal R^n=\bigcup_{j=1}^m\mathcal R_{j,j+1}^n
\end{equation}
and
\begin{equation}\label{sp25c}
\mathcal G^n=\bigcup_{j=1}^m\mathcal G_j^n.
\end{equation}
By definition,
the $j$-th {\it petal}, $\mathcal P_j^n$, of the rosette $\mathcal H^n$ is the region bounded 
by the rays $\mathcal R_{j,j+1}^n$, $\mathcal R_{j-1,j}^n$, and the arc $\mathcal G_j^n$.

{\it Construction of the spurious zeros.} Consider the function,
\begin{equation}\label{sp26}
h_j^n(\z)=\frac{e \z e^{-\la_j \z}}{\left[c_j\sqrt{2\pi n}\,(1-\z)\right]^{\frac{1}{n}}}.
\end{equation}
It maps the arc $\mathcal G_j^n$ into the unit circle. Define the points $\z_q(j,n)$ as the
preimages of the points $\om_q$ on the arc $\mathcal G_j^n$,
\begin{equation}\label{sp27}
\z_q(j,n)=(h_j^n)^{-1}(\om_q),
\qquad \z_q(j,n)\in \mathcal G_j^n. 
\end{equation}
Let
\begin{equation}\label{sp27a}
d_{jn}(\z)=n\,\min\left\{\,\dist(\z,\z_c^n(j,j+1)),\,
\dist(\z,\z_c^n(j-1,j))\,\right\}.
\end{equation}

\begin{theo} \label {spurious_zeros_not_1}
{\rm (Spurious zeros, $j\not=1$.)} There exists $R_0>0$ such that 
if $j\not=1$ then
for any $\z_q(j,n)\in \mathcal G_j^n$ such that
\begin{equation}\label{sp28}
d_{jn}(\z_q(j,n))>R_0,
\end{equation}
there exists a unique simple zero $\z_k(n)$ of $f_{n-1}(n\z)$ such that
\begin{equation}\label{sp29}
\z_k(n)=\z_q(j,n)+O(n^{-1}e^{-d_{jn}(\z_q(j,n))}+n^{-2}),\qquad n\to\infty.
\end{equation}
\end{theo}

\begin{proof} 
By using (\ref{sp26}), we write equation (\ref{sp4}),
which is equivalent to the equation $f_{n-1}(z)=0$, as
\begin{equation}\label{sp29a}
\om_q(1+O(e^{-d_j(z)}))^{\frac{1}{n}}=h_j^n(\z)(1+O(n^{-2})),
\end{equation}
or as
\begin{equation}\label{sp30}
h_j^n(\z)=\om_q+O(n^{-1}e^{-d_{jn}(\z)}+n^{-2}).
\end{equation}
If $j\not=1$, then the generalized Szeg\"o curve, $\Gamma(|\la_j|)$,
lies strictly inside of the unit circle, hence the arc $\mathcal G_j$ does. 
Observe that in a neighborhood of $\mathcal G_j$,
\begin{equation}\label{sp31}
\lim_{n\to\infty}h_j^n(\z)=h_j(\z)\equiv e\z e^{-\la_j\z},
\end{equation}
and $h'_j(\z)\not=0$, hence $(h_j^n)'(\z)\to h'_j(\z)$ and
$|(h^n_j)'(\z)|>\ep>0$ for large $n$. Since $|\om_q-\om_{q+1}|=\frac{2\pi}{n}+O(n^{-2})$,
we obtain, by Proposition \ref{hurwitz}, that if $d_{jn}(\z_q(j,n))$ is big then there is
a simple root $\z_k(n)$ of equation (\ref{sp30}) such that (\ref{sp29}) holds, QED.
\end{proof}

For $j=1$ the Szeg\"o curve, $\Gamma(1)$, is not strictly inside of the unit disk,
because it contains the point $\z=1$. Let us fix some $\rho<1$ sufficiently
close to 1 and  partition the zeros $\z_k(n)$ of $f_{n-1}(n\z)$
into two groups: $|\z_k(n)|\le \rho$ and $|\z_k(n)|>\rho$. For the first
group we will prove formula (\ref{sp29}) with $j=1$. For the second group
we will consider another approximation to $\z_k(n)$. Let $\z_k^0(n)$ be the zeros
of the section $s_{n-1}(n\z)$ of $e^{n\z}$, so that
\begin{equation}\label{sp34}
s_{n-1}(z)=\sum_{k=0}^{n-1}\frac{z^k}{k!}\,.
\end{equation}
We will prove for the second group, that $\z_k(n)$ is well approximated by $\z_k^0(n)$.
The asymptotics $\z_k^0(n)$ is well-known for $|\z_k^0(n)|<1-\ep$, 
$\ep>0$, and for $|\z_k^0(n)-1|\le \frac{C}{\sqrt n}\,$. In Appendices \ref{uniform} and \ref{s_n} below
we derive uniform asymptotics of $\z_k^0(n)$ in the disk $D(1,\de)$, $\de>0$.

\begin{theo} \label {spurious_zeros_1}
{\rm (Spurious zeros, $j=1$.)} There exist $1>\rho_0>0$ and  $R_0>0$ such that
for any $\rho\in(\rho_0,1)$ the following is true: 
For any $\z_q(1,n)\in \mathcal G_1^n$ such that
\begin{equation}\label{sp35}
d_{1n}(\z_q(1,n))>R_0,
\end{equation}
and
\begin{equation}\label{sp36}
|\z_q(1,n)|<\rho,
\end{equation}
there exists a unique simple zero $\z_k(n)$ of $f_{n-1}(n\z)$ such that
\begin{equation}\label{sp37}
\z_k(n)=\z_q(1,n)+O(n^{-1}e^{-d_{1n}(\z_q(1,n))}+n^{-2}),\qquad n\to\infty.
\end{equation}
In addition, for any zero $\z_k^0(n)$ of $s_{n-1}(n\z)$ such that
\begin{equation}\label{sp38}
|\z_k^0(n)|>\rho,
\end{equation}
there exists a unique simple zero $\z_k(n)$ of $f_{n-1}(nz)$ such that
\begin{equation}\label{sp39}
\z_k(n)=\z_k^0(n)+O(e^{-cn}),\quad n\to\infty;\qquad
c>0.
\end{equation}
\end{theo}

\begin{proof} 
The first part of the theorem, about (\ref{sp37}), is proved in the same way as
Theorem \ref{spurious_zeros_not_1}, and we omit the proof. For the second part, 
observe that for any $\de>0$ there exists $\rho_0<1$ such that for any $\rho\in(\rho_0,1)$,
condition (\ref{sp38}) implies that
\begin{equation}\label{sp39a}
|\z_k^0(n)-1|<\de.
\end{equation}
Let us  write the equation,
\begin{equation}\label{sp40}
f_{n-1}(z)=s_{n-1}(z)+f_{n-1}^1(z)=0,\quad
f_{n-1}^1(z)=\sum_{k=0}^{n-1}\frac{(c_2\la_2^k+\ldots c_M\la_M^k)z^k}{k!}\,,
\end{equation}
in the form
\begin{equation}\label{sp41}
g_n(\z)\equiv \frac{s_{n-1}(n\z)}{e^{n\z}}=-e^{-n\z}f_{n-1}^1(n\z),
\qquad z=n\z.
\end{equation}
Observe that if $\ep>0$ is sufficiently small then there exists $c>0$
such that for any $\z$ in the disk $\{|\z-1|\le\ep\}$,
\begin{equation}\label{sp42}
e^{-n\z}f_{n-1}^1(z)=O(e^{-cn}),\qquad n\to\infty.
\end{equation}
The zeros $\z_k^0(n)$ solve the equation $g_n(\z_k^0(n))=0$,
and inequality (\ref{sn68}) in Appendix \ref{s_n} below implies that
\begin{equation}\label{sp43}
|g_n(\z)|\ge |\z-\z_k^0(n)|,\quad {\rm if}\quad
|\z-\z_k^0(n)|\le cn^{-1}.
\end{equation}
Therefore, by Proposition \ref{hurwitz}, for any zero $\z_k^0(n)$ of $s_{n-1}(n\z)$ there
exists a unique simple zero $\z_k(n)$ of $f_{n-1}(n\z)$ such that (\ref{sp39}) holds. Theorem
\ref{spurious_zeros_1} is proved.
\end{proof}

The zeros $z_k(n)$ described in Theorems \ref{spurious_zeros_not_1} and \ref{spurious_zeros_1}
are called the {\it spurious zeros}  of the section $f_{n-1}(z)$. For these zeros,
$\z_k(n)=z_k(n)/n$ lies
in a small neighborhood of the curve $\mathcal G^n$.

\section{Transitional zeros}\label{TRZ}

For a given $j=1,\ldots,m$, the $j$-th set, $\mathcal T^n_j$, of the transitional zeros
of $f_{n-1}$
is located in a neighborhood of the point $n\z_c^n(j,j+1)$. Let us first describe
$\mathcal T^n_j$ informally. We set
\begin{equation}\label{tr1}
z=n\z_c^n(j,j+1)+w
\end{equation}
and substitute this into equation (\ref{ms23}), which is equivalent to $f_{n-1}(z)=0$. This gives that
\begin{equation}\label{tr2}
\cos u+O(e^{-c\Re u})=\frac{e^n(\z_c^n+\frac{w}{n})^ne^{-\frac{(\la_{j+1}+\la_j)(n\z_c^n+w)}{2}}}
{2\sqrt{c_jc_{j+1}}\,\sqrt{2\pi n}\,(1-\z_c^n+\frac{w}{n})}(1+O(n^{-1})).
\end{equation}
We will assume that $w=O(n^{\frac{1}{3}})$ as $n\to\infty$. Then
\begin{equation}\label{tr3}
\left(\z_c^n+\frac{w}{n}\right)^n=(\z_c^n)^ne^{(\z_c^n)^{-1}w}(1+O(n^{-\frac{1}{3}})),
\end{equation}
and (\ref{tr2}) reduces to the equation,
\begin{equation}\label{tr4}
\cos u+O(e^{-c\Re u})=A_ne^{\left(\z_c^{-1}-\frac{\la_{j+1}+\la_j}{2}\right)w}
(1+O(n^{-\frac{1}{3}})),
\end{equation}
where
\begin{equation}\label{tr5}
A_n=\frac{e^n(\z_c^n)^n e^{-\frac{(\la_{j+1}+\la_j)n\z_c^n}{2}}}
{2\sqrt{c_jc_{j+1}}\,\sqrt{2\pi n}\,(1-\z_c^n)}.
\end{equation}
By (\ref{ms24}), 
\begin{equation}\label{tr6}
|A_n|=1,
\end{equation}
and by (\ref{dom12a}),
\begin{equation}\label{tr7}
e^{iu}=e^{\frac{(\la_{j+1}-\la_j)z}{2}}\sqrt{\frac{c_{j+1}}{c_j}}.
\end{equation}
By substituting (\ref{tr1}), we obtain that
\begin{equation}\label{tr8}
e^{iu}=e^{\frac{(\la_{j+1}-\la_j)n\z_c^n}{2}}e^{\frac{(\la_{j+1}-\la_j)w}{2}}\sqrt{\frac{c_{j+1}}{c_j}}.
\end{equation}
Observe that
\begin{equation}\label{tr9}
(\la_{j+1}-\la_j)\z_c^n=(\la_{j+1}-\la_j) r_c^n e^{i\theta_{j,j+1}}=i r_c^n|\la_{j+1}-\la_j|,
\end{equation}
hence
\begin{equation}\label{tr10}
e^{iu}=B_n e^{\frac{(\la_{j+1}-\la_j)w}{2}},
\end{equation}
where
\begin{equation}\label{tr11}
B_n=e^{\frac{inr_c^n|\la_{j+1}-\la_j|}{2}}\sqrt{\frac{c_{j+1}}{c_j}}.
\end{equation}
We have that
\begin{equation}\label{tr12}
|B_n|=\sqrt{\left|\frac{c_{j+1}}{c_j}\right|}.
\end{equation}
Thus, equation (\ref{tr4}) reduces to
\begin{equation}\label{tr13}
B_n e^{\frac{(\la_{j+1}-\la_j)w}{2}}+B_n^{-1} e^{\frac{(\la_j-\la_{j+1})w}{2}}
=2A_ne^{\left(\z_c^{-1}-\frac{\la_{j+1}+\la_j}{2}\right)w}+O(n^{-\frac{1}{3}}),
\end{equation}
or
\begin{equation}\label{tr14}
B_n e^{\la_{j+1}w}+B_n^{-1} e^{\la_j w}
-2A_ne^{\z_c^{-1}w}=O(n^{-\frac{1}{3}}).
\end{equation}
The expression on the left is a  three term exponential sum with the exponents
$\la_{j+1}$, $\la_j$, and $\z_c^{-1}$. Observe that 
\[
\left|\z_c^{-1}\right|=r_c^{-1}>1,
\]
and the vector $\z_c^{-1}$ is orthogonal to the one $(\la_{j+1}-\la_j)$.
Since $|\la_j|,\;|\la_{j+1}|\le 1$, this implies that the numbers 
$\la_{j+1}$, $\la_j$, and $\z_c^{-1}$ do not lie on the same line.
Now we can formulate the asymptotic formula for the transitional zeros.

\begin{theo} \label{transitional_zeros}
{\rm (Transitional zeros.)} There is a one-to-one correspondence between the zeros $w_k(n)$ of the
exponential sum,
\begin{equation}\label{tr15} 
g_n(w)=B_n e^{\la_{j+1}w}+B_n^{-1} e^{\la_j w}
-2A_ne^{\z_c^{-1}w},
\end{equation}
in the disk $D(0,n^{\frac{1}{3}})$ and the zeros $z_k(n)$ of $f_{n-1}(z)$ 
in the disk $D(n\z_c^n(j,j+1),n^{\frac{1}{3}})$, 
such that
\begin{equation}\label{tr16}
z_k(n)=n\z_c^n(j,j+1)+w_k(n)+O(n^{-\frac{1}{6}}).
\end{equation}
\end{theo}

\begin{proof}  
It follows from (\ref{tr6}) and (\ref{tr12}) that there exists $R>0$, independent of $n$,
such that all the zeros of $g_n(w)$ with $|w|>R$ are simple and belong 
to the main series, see Section \ref{ZES}. For them equation (\ref{tr14}) implies (\ref{tr16}),
with even a better error term, $O(n^{-\frac{1}{3}})$. It remains to consider zeros $|w_k(n)|\le R$.

The zeros of $g_n(w)$ are at most double, since the Vandermonde determinant is nonzero.
Therefore, there exists $\ep>0$, independent of $n$, such that for any three zeros, $w_k(n)$, $w_l(n)$,
$w_m(n)$, of $g_n(w)$,
\begin{equation}\label{tr16a}
\max\,\{\,|w_k(n)-w_l(n)|,\;|w_k(n)-w_m(n)|,\;|w_m(n)-w_l(n)|\,\}>\ep.
\end{equation}
This implies that there exists $\exists\,\de>0$ such that for any $w_k(n)\in D(0,R)$,
there exists $0.1\ep<r<0.2\ep$ such that 
\begin{equation}\label{tr16b}
|g_n(w)|>\de,\qquad \forall\, |w-w_k(n)|=r.
\end{equation}
Equation (\ref{tr14}) implies now that $\exists\,N>0$ such that $\forall\,n>N$,
there exists a zero $z_k(n)$ of $f_{n-1}$ 
which satisfies (\ref{tr16}). Relation (\ref{tr16a}) enables us to make the
correspondence $w_k(n)\to z_k(n)$ one-to-one. Theorem \ref{transitional_zeros}
is proved.
\end{proof}

We call the zeros $z_k(n)$ satisfying (\ref{tr16}), the {\it transitional zeros in 
the neighborhood of the point $n\z_c^n(j,j+1)$}. The set of these zeros 
is denoted by $\mathcal T^n_j$. The set of all the transitional 
zeros is
\begin{equation}\label{tr17}
\mathcal T^n=\bigcup_{j=1}^m \mathcal T^n_j.
\end{equation}
The set of transitional zeros overlaps with both the zeros of the main series
and the spurious zeros.

\section{Completeness of Zeros}

\begin{theo} \label {completeness}
{\rm (Completeness.)} There exists $N>0$ such that for any $n>N$,
any zero $z_k(n)$ of $f_{n-1}$ belongs to one of the following four categories: 
finite zeros, zeros of the main series, spurious zeros, and transitional zeros.
\end{theo}

\begin{proof} 
We will consider zeros in different regions on the complex plane.
As usual, $z_k(n)=n\z_k(n)$.

{\it Region 1, $z_k(n)\in D(0,R_0)$.}
 The completeness follows from Proposition \ref{finite_zeros}.

{\it Region 2, $\z_k(n)\in \C\setminus[D(0,\rho)\cup D(1,\de)]$, $0<\rho<1$, $0<\de$.}

\begin{lem} \label{outside}
{\rm (Absence of zeros outside of $D(0,\rho)\cup D(1,\de)$.)}  For any $\de>0$
there exist $1>\rho>0$ and $N>0$ such that for any $n>N$,
 $f_{n-1}(n\z)\not=0$ if $\z\not\in D(0,\rho)\cup D(1,\de)$.
\end{lem}

\begin{proof} 
Fix any $\de>0$.
The function $s_{n-1}(n\z)$ is a polynomial of degree $(n-1)$ and
for large $\z$ the leading term of this polynomial dominates,
\begin{equation}\label{com2}
s_{n-1}(n\z)=\frac{n^{n-1}\z^{n-1}}{(n-1)!}(1+o(1)),\quad n\to\infty.
\end{equation}
On the other hand, for small $\z$,
\begin{equation}\label{com3}
s_{n-1}(n\z)=e^{n\z}(1+o(1)),\quad n\to\infty.
\end{equation}
The transition from one asymptotics to another occurs 
in a neighborhood of the Szeg\"o curve.
Let $\Ga_\ep$, $\ep>0$, be the
$\ep$-neighborhood of the Szeg\"o curve $\Ga$, and let $D^{\rm int}_\ep$ and
$D^{\rm out}_\ep$  be the two connected components of $\C\setminus \Ga_\ep$,
interior and exterior. Then, as shown by Szeg\"o \cite{Sze}, 
asymptotics (\ref{com2}) holds in $D^{\rm int}_\ep$, while asymptotics 
(\ref{com3}) holds in $D^{\rm out}_\ep$, with a uniform with respect to $z$
estimate of the error terms $o(1)$. It follows from (\ref{com2}), that
for any $\ep_0>0$ there exists $N>0$ such that
\begin{equation}\label{com4}
e^{n(1+\ep_0)}|\z|^{n-1}\ge |s_{n-1}(n\z)|\ge e^{n(1-\ep_0)}|\z|^{n-1},
\quad \z\in D^{\rm out}_\ep,\quad n>N.
\end{equation}
It is obvious that for all $\z$,
\begin{equation}\label{com5}
|s_{n-1}(n\z)|\le \sum_{k=0}^\infty \frac{|n\z|^k}{k!}=e^{n|\z|}.
\end{equation}
For the given $\de>0$, let us choose $\ep>0$, $\ep_0>0$, $\ep_1>0$,
and $1>\rho>0$ such that the following three conditions are satisfied:
\begin{enumerate}
\item
\begin{equation}\label{com5_1}
\Ga_\ep\subset D(0,\rho)\cup D(1,\de)\,;
\end{equation}
\item
\begin{equation}\label{com5_2}
|\la_j|(1+\ep_1)<1-\ep_0-\ep+\ln\rho\,,
\qquad j=2,\ldots,M\,;
\end{equation}
and
\item
\begin{equation}\label{com5_3}
|\la_j|<e^{-2\ep_0-\ep},\qquad j=2,\ldots,M\,.
\end{equation}
\end{enumerate}
We claim that then  $\exists\,N>0$ such that for $j=2,\ldots,M$,
\begin{equation}\label{com5a}
|s_{n-1}(n\la_j\z)|\le e^{-\ep n}|s_{n-1}(n\z)|,
\quad \z\not\in D(0,\rho)\cup D(1,\de),
\quad n>N.
\end{equation}
Indeed, consider two cases: (1) $\rho\le |\z|\le 1+\ep_1,$ 
and (2) $|\z|> 1+\ep_1$. In case (1), by (\ref{com4}), (\ref{com5}), 
and (\ref{com5_2}),
\begin{equation}\label{com6}
|s_{n-1}(n\la_j\z)|\le e^{n|\la_j\z|}\le e^{n(1-\ep_0-\ep)}\rho^{n-1}
\le e^{n(1-\ep_0-\ep)}|\z|^{n-1}\le e^{-\ep n}|s_{n-1}(n\z)|,
\quad n>N,
\end{equation}
and in case (2), by (\ref{com4}) and (\ref{com5_3}),
\begin{equation}\label{com8}
|s_{n-1}(n\la_j\z)|\le e^{n(1+\ep_0)}|\la_j\z|^{n-1}
\le e^{n(1-\ep_0-\ep)}|\z|^{n-1}
\le e^{-\ep n}|s_{n-1}(n\z)|,
\quad n>N.
\end{equation}
This proves (\ref{com5a}). From (\ref{com5a}) we obtain that there exists $N>0$ such that
\begin{equation}\label{com9}
|f^1_{n-1}(n\z)|\le Me^{-\ep n}|s_{n-1}(n\z)|,\quad \z\not\in D(0,\rho)\cup D(1,\de),
\quad n>N,
\end{equation}
hence $f_{n-1}(n\z)=s_{n-1}(n\z)+f_{n-1}^1(n\z)\not=0$, if $N$ is big enough, QED.
\end{proof}

{\it Region 3, $z_k(n)\in \mathcal S_{j,j+1}(r_0,R_0)\cap D(0,n\rho)$.} We go back
to Section \ref{ZMS}. In the disk $\z\in D(0,\rho)$, function (\ref{ms21d}) 
is well approximated by (\ref{ms21e}), hence the equation $f_{n-1}(n\z)=0$ reduces
to (\ref{ms23}). Therefore,  if 
$z_k(n)\in \mathcal S_{j,j+1}(r_0,R_0)\cap D(0,n\rho)$
and $|z|>nr_c^n+0.5n^{\frac{1}{3}}$, then equation $f_{n-1}(n\z)=0$ has no zeros
for large $n$, 
because the absolute value of the right hand side in (\ref{ms23}) approaches infinity,
as $n\to\infty$, while the left hand side remains bounded. This proves
that the only zeros of $f_{n-1}$ in $\mathcal S_{j,j+1}(r_0,R_0)\cap D(0,n\rho)$
are the zeros of the main series and transitional zeros.

{\it Region 4, $z_k(n)\in \mathcal U_{j}(r_0,R_0)\cap D(0,n\rho)$.} We go back
to Section \ref{SPZ}. In the disk $\z\in D(0,\rho)$, function (\ref{ms21d}) 
is well approximated by (\ref{ms21e}), hence the equation $f_{n-1}(n\z)=0$ reduces
to (\ref{sp30}). This implies that all the zeros of $f_{n-1}$ in
$\mathcal S_{j,j+1}(r_0,R_0)\cap D(0,n\rho)$ are either spurious or
transitional.  

{\it Region 5, $\z_k(n)\in D(1,\de)$.} According to (\ref{sp41}), (\ref{sp42}),
the equation $f_{n-1}=0$ reduces to the one,
\begin{equation}\label{com10}
\frac{s_{n-1}(n\z)}{e^{n\z}}=O(e^{-cn}).
\end{equation}
Asymptotics (\ref{sn38}), in Appendix \ref{s_n}
below, proves that all the zeros of the latter equation are 
located near the zeros of $s_{n-1}$, which implies that all these zeros are spurious,
described by (\ref{sp39}).

This ends the proof of Theorem \ref{completeness}.
\end{proof}

\section{Limiting Distribution of Zeros on the Rosette}

It follows from Theorems \ref{main_series}, \ref{spurious_zeros_not_1},
and \ref{spurious_zeros_1} that as $n\to\infty$,
the normalized zeros $\z_k=\frac{z_k}{n}$ of the section $f_{n-1}$
approach the rosette $\mathcal H$, and the $\de$-function measure of zeros,
\begin{equation}\label{ld1}
d\mu_{n-1}=\frac{1}{n-1}\sum_{k=1}^{n-1}\de(\z-\z_k)\,d\z,
\end{equation}
weakly converges to a probability measure on $\mathcal H$,
\begin{equation}\label{ld2}
\lim_{n\to\infty}\mu_{n-1}=\mu_{\mathcal H},
\end{equation}
such that for any continuous test function $\f(\z)$,
\begin{equation}\label{ld3}
\lim_{n\to\infty}\int \f(\z) d\mu_{n-1}
=\int \f(\z) d\mu_{\mathcal H}=\int_{\mathcal H} \f(\z)p(\z)|d\z|,
\end{equation}
where $p(\z)\ge 0$ is a density function on $\mathcal H$. The above theorems
give the following description of the density $p(\z)$.

\begin{theo} \label{density}
On the ray $\mathcal R_{j,j+1}$, $p(\z)$ is constant,
\begin{equation}\label{ld4}
p(\z)=\frac{|\la_{j+1}-\la_j|}{2\pi}\,,\quad \z\in \mathcal R_{j,j+1},
\end{equation}
$j=1,\ldots,m$. On the curve $\mathcal G_j$,
\begin{equation}\label{ld5}
p(\z)=\frac{|h'_j(\z)|}{2\pi}\,,\quad \z\in \mathcal G_j,
\end{equation}
where
\begin{equation}\label{ld6}
h_j(\z)=\z e^{(1-\la_j)\z}.
\end{equation}
\end{theo}

\begin{proof} 
Ray $\mathcal R_{j,j+1}$. 
By (\ref{ms27}) the scaled zeros $\z_k(n)$ of
$f_{n-1}$ are close to the scaled zeros $\z_k$
of $f$, so that
\begin{equation}\label{ld7}
\z_k(n)-\z_k=O(n^{-1}e^{-\ga n(r_c^n-|\z|_k)}).
\end{equation}
On the other hand, by (\ref{dom17}) for $z_k\in S_{j,j+1}(r_0,R_0)$,
\begin{equation}\label{ld8}
\z_k-n^{-1}\left(\al_{j,j+1}+\frac{2\pi li}{\la_{j+1}-\la_j}\right)
=O(n^{-1}e^{-cl}).
\end{equation}
This implies that $\z_k(n)$ are close to the points
of the lattice 
\begin{equation}\label{ld9}
\mathcal L_{j,j+1}=\left\{z=n^{-1}\left(\al_{j,j+1}
+\frac{2\pi li}{\la_{j+1}-\la_j}\right),\quad l\in\Z\right\},
\end{equation}
hence (\ref{ld4}) follows.

Curve $\mathcal G_j$, $j\not=1$. From (\ref{sp29}) we obtain that
\begin{equation}\label{ld10}
\begin{aligned}
\z_k(n)&=(h_j^n)^{-1}(\om_q)+O(n^{-1}e^{-d_{jn}(\z_q(j,n))}+n^{-2})\\
&=(h_j)^{-1}(\om_q)+O(n^{-1}e^{-d_{jn}(\z_q(j,n))}+n^{-2}\ln n),
\end{aligned}
\end{equation}
hence
\begin{equation}\label{ld11}
h_j(\z_k(n))
=\om_q+O(n^{-1}e^{-d_{jn}(\z_q(j,n))}+n^{-2}\ln n),
\end{equation}
If $\z=\z_k(n)$ and $\z^0$ is $\z_k(n)$ which corresponds to $\om_{q+1}$,
then 
\begin{equation}\label{ld12}
\om_{q+1}-\om_q=h'_j(\z)(\z^0-\z)+O(n^{-1}e^{-d_{jn}(\z_q(j,n))}+n^{-2}\ln n),
\end{equation}
which implies (\ref{ld5}). This proves Theorem \ref{density}.
\end{proof}

\section{ Beyond Conditions P and Q}

\subsection{Beyond Condition P}

Condition P means that there is no $\la_j$, $j=m+1,\ldots,M$, on the sides
of the polygon $P_m$. Suppose that this condition does not hold and there are
some $\la_k$'s on the side $[\la_j,\la_{j+1}]$. Denote
\begin{equation}\label{bd1}
\sigma_j=\{ \la_k:\; \la_k\in [\la_j,\la_{j+1}]\}.
\end{equation}
Observe that $\sigma_j$ includes $\la_j$ and $\la_{j+1}$. In this case,
instead of two-term equation (\ref{dom12}), we consider the multiterm equation,
\begin{equation}\label{bd2}
\sum_{k:\;\la_k\in\sg_j} c_k e^{\la_k z}=0.
\end{equation}
With the help of substitution (\ref{dom12a}) we reduce it to the equation
\begin{equation}\label{bd3}
\sum_{k:\;\la_k\in\sg_j} \tilde c_k e^{i y_k u}=0,\qquad -1\le y_k\le 1.
\end{equation}
The function on the left is quasiperiodic.
Its zeros are concentrated in a finite strip $\{ u:\; |\Im u|<A\}$.
The distribution of zeros of quasiperiodic exponential sums 
was studied in the work of Soprunova \cite{Sop}. 
It was shown that the zeros also have a property of quasiperiodicity,
and its average number is the same as for the function $\cos u$. 

The main results of the present paper, concerning the distribution of zeros
of the exponential sum and its sections, can be extended, with proper 
modifications, to the case when Condition P does not hold, but
it requires some additional considerations for the zeros
of the main series.

\subsection{Beyond Condition Q} 

If  Condition Q violates then there are several maximal $|\la_j|$'s. As an 
example, consider the symmetric sum,
\begin{equation}\label{bd4}
f(z)=\frac{1}{m}\sum_{j=0}^{m-1}  e^{\om^j z}=0,\qquad \om=e^{\frac{2\pi i}{m}}\,.
\end{equation}
In this case,
\begin{equation}\label{bd5}
f_n(z)=\sum_{k:\; 0\le mk\le n} \frac{z^{mk}}{(mk)!}\,.
\end{equation}
The rosette is symmetric and all the results are extended to this case. 
Figure 5 depicts zeros of the $n=200$ section of exponential sum (\ref{bd4})
for $m=4$. The spurious zeros are described in the symmetric case
by parts of the original Szeg\"o curve,
$\Ga(1)$.

\begin{center}
\begin{figure}[h]\label{figure5}
\begin{center}
\scalebox{0.5}{\includegraphics{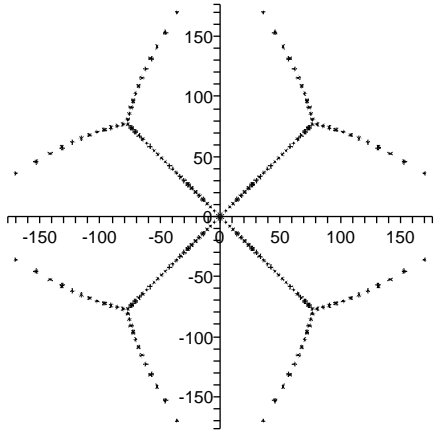}}
\end{center}
   \caption{The zeros of the $n=200$ section of exponential sum (\ref{bd4}) for $m=4$.}
   \end{figure}
\end{center}

All the results are extended also to a slightly more general case of
\begin{equation}\label{bd4a}
f(z)=\sum_{j=0}^{m-1} c_j e^{a\om^j z}=0,\qquad \om=e^{\frac{2\pi i}{m}}\,,
\end{equation}
where $c_j\not=0$, $j=0,\ldots,m-1,$ and $a\not=0$.
For $m=2$ this includes the sine and cosine functions. 

Consider now the exponential sum
\begin{equation}\label{bd6}
f(z)=\sum_{j=1}^{m}  e^{\la_j z},
\end{equation}
where
\begin{equation}\label{bd7}
\la_j=e^{i\f_j},\quad \f_j\in\R.
\end{equation}
Then
\begin{equation}\label{bd8}
f_n(z)=\sum_{k=0}^{n}  a_kz^k,
\end{equation}
where
\begin{equation}\label{bd9}
a_k=\frac{e^{ik\f_1}+\ldots+e^{ik\f_m}}{k!}.
\end{equation}
In this case the asymptotic behavior of the coefficients $a_k$ 
depends on the arithmetic properties of the numbers $\f_j$,
and the asymptotic behavior of the zeros $z_k$ of $f_n$ can be
rather complicated.

\begin{appendix}

\section {Uniform asymptotics of the zeros of $s_{n-1}(n\z)$
in the disk $D(1,\de)$}  \label{uniform}

We write the equation
\begin{equation}\label{exp1}
s_{n-1}(n\z)\equiv \sum_{k=0}^{n-1}\frac{(n\z)^k}{k!}=0
\end{equation}
as
\begin{equation}\label{exp2}
e^{n\z}=\frac{(n\z)^n}{n!}\sum_{k=0}^\infty \frac{n! (n\z)^k}{(n+k)!}\,,
\end{equation}
or
\begin{equation}\label{exp3}
\frac{e^{-n\z}n^n \z^n}{n!}\sum_{k=0}^\infty \frac{n! n^k\z^k}{(n+k)!}=1.
\end{equation}
We will assume that
\begin{equation}\label{exp4}
|\z|\le 1.
\end{equation}
With the help of the Stirling formula we obtain that
\begin{equation}\label{exp5}
\frac{n! n^k}{(n+k)!}
=\exp\left[ k-\left(n+k+\frac{1}{2}\right)\ln\left(1+\frac{k}{n}\right)
+\frac{\theta_n}{12n}-\frac{\theta_{n+k}}{12(n+k)}\right]\,.
\end{equation}
Since
\begin{equation}\label{exp5a}
\ln(1+x)=x-\frac{x^2}{2}+O(x^3)\,,\quad x\to 0,
\end{equation}
this gives that as $n\to\infty$,
\begin{equation}\label{exp6}
\frac{n! n^k}{(n+k)!}=e^{-\frac{k^2}{2n}}
\left(1+O\left(\frac{k^3}{n^2}+\frac{k}{n}\right)\right),\quad 0\le k\le n^{0.6}.
\end{equation}
Also,
\begin{equation}\label{exp6a}
\frac{n! n^k}{(n+k)!}=
\left\{ 
\begin{aligned}
&O(e^{-n^{0.1}}),\quad  n^{0.6}\le k\le n,\\
&O(2^{n-k}e^{-n^{0.1}}),\quad  n\le k.
\end{aligned}
\right.
\end{equation}
Therefore,
\begin{equation}\label{exp7}
\sum_{k=0}^\infty \frac{n! n^k\z^k}{(n+k)!}=\sum_{k=0}^\infty e^{-\frac{k^2}{2n}}\z^k+O(1),
\quad |\z|\le 1,
\end{equation}
and equation (\ref{exp3}) reduces to
\begin{equation}\label{exp7a}
\frac{e^{-n\z}n^n \z^n}{n!}\left[\sum_{k=0}^\infty e^{-\frac{k^2}{2n}}\z^k+O(1)\right]=1,
\qquad |\z|\le 1.
\end{equation}
Let
\begin{equation}\label{exp8}
\z=e^{\tau n^{-1/2}},\quad \Re\tau\le 0.
\end{equation}
Then
\begin{equation}\label{exp9}
\sum_{k=0}^\infty e^{-\frac{k^2}{2n}}\z^k
=\sum_{k=0}^\infty e^{-\frac{k^2}{2n}+\tau kn^{-1/2}}
=n^{\frac{1}{2}}\int_0^\infty e^{-\frac{x^2}{2}+\tau x}dx+O(|\tau|+1),
\end{equation}
by the Euler-Maclaurin integration formula. To justify the error term, observe that 
\begin{equation}\label{exp9a}
\left|e^{-\frac{k^2}{2n}+\tau kn^{-1/2}}-\int_k^{k+1}e^{-\frac{x^2}{2n}}e^{\tau xn^{-1/2}}dx\right|
\le \int_k^{k+1}\left(\frac{x}{n}+|\tau| n^{-1/2}\right)|e^{-\frac{x^2}{2n}+\tau xn^{-1/2}}|dx
\end{equation}
hence
\begin{equation}\label{exp9b}
\begin{aligned}
\left|\sum_{k=0}^\infty e^{-\frac{k^2}{2n}+\tau kn^{-1/2}}
-\int_0^{\infty}e^{-\frac{x^2}{2n}+\tau xn^{-1/2}}dx\right|
&\le \int_0^\infty\left(\frac{x}{n}+|\tau| n^{-1/2}\right)|e^{-\frac{x^2}{2n}+\tau xn^{-1/2}}|dx\\
&\le C(1+|\tau|),
\end{aligned}
\end{equation}
which implies (\ref{exp9}).
From (\ref{exp7}) and (\ref{exp9}),
\begin{equation}\label{exp10}
\sum_{k=0}^\infty \frac{n! n^k\z^k}{(n+k)!}
=n^{\frac{1}{2}}\int_0^\infty e^{-\frac{x^2}{2}+\tau x}dx+O(|\tau|+1),
\end{equation}
Equation (\ref{exp3}) reduces to
\begin{equation}\label{exp11}
\frac{e^{-n\z}n^n \z^n}{n!}
\left[n^{\frac{1}{2}}\int_0^\infty e^{-\frac{x^2}{2}+\tau x}dx+O(|\tau|+1)\right]=1.
\end{equation}
By applying the Stirling formula, we obtain that
\begin{equation}\label{exp12}
e^{n(1-\z)} \z^n 
\left[\frac{1}{\sqrt{2\pi}}
\int_0^\infty e^{-\frac{x^2}{2}+\tau x}dx
+O\left((|\tau|+1)n^{-\frac{1}{2}}\right)\right]=1.
\end{equation}
Let us fix any big number $M>0$ and any small number $\ep>0$,
and consider the three (partially overlapping) cases: 
\begin{enumerate}
\item
$0\le |\tau|\le M$,
\item
$M\le |\tau|\le n^{\frac{1}{6}-\ep}$, $\Re\tau<0$,
\item
$ \Re\tau\le -n^{\ep}$.
\end{enumerate}

{\it Case (1), $0\le |\tau|\le M$.} We assumed $\Re\tau\le 0$, but if $|\tau|\le M$,
equations (\ref{exp7a}) and (\ref{exp9}) hold without this restriction. In the case
under consideration,
\begin{equation}\label{exp13}
\z=e^{\tau n^{-\frac{1}{2}}}=1+\tau n^{-\frac{1}{2}}+\frac{\tau^2 n^{-1}}{2}+O(n^{-\frac{3}{2}}),
\end{equation}
hence
\begin{equation}\label{exp14}
e^{n(1-\z)} \z^n =e^{-\frac{\tau^2}{2}}(1+O(n^{-\frac{1}{2}})),
\end{equation}
hence equation (\ref{exp12}) reduces to
\begin{equation}\label{exp15}
\frac{1}{\sqrt{2\pi}}
\int_0^\infty e^{-\frac{(x-\tau)^2}{2}}dx+O(n^{-\frac{1}{2}})=1,
\end{equation}
or
\begin{equation}\label{exp16}
\frac{1}{\sqrt{2\pi}}
\int_{\tau}^\infty e^{-\frac{x^2}{2}}dx+O(n^{-\frac{1}{2}})=0,
\end{equation} 
which is the Newman-Rivlin equation
\cite{NR}, with an error term of the order
of $n^{-\frac{1}{2}}$.
All zeros of the function, 
\begin{equation}\label{exp17}
E(\tau)=\frac{1}{\sqrt{2\pi}}
\int_{\tau}^\infty e^{-\frac{x^2}{2}}dx,
\qquad E(\tau)=\frac{1}{2}\,\Erfc\left(\frac{\tau}{\sqrt 2}\right),
\end{equation}
are simple and they lie in the left half-plane, $\{z:\;\Re z<0\}$,
see \cite{FCC}.
If we enumerate the zeros $\tau_q$ in the second quadrant by $|\tau_q|$, then
\[
\tau_1=-1.915990857\ldots+i2.816359418\ldots,
\]
 and
\begin{equation}\label{exp17a}
\tau_q=2\sqrt{\pi q}e^{\frac{3\pi i}{4}}+
\frac{1}{4\sqrt{\pi q}}\ln(8\pi q) e^{\frac{\pi i}{4}}+
O\left( q^{-\frac{1}{2}}\right),\qquad
q\to\infty.
\end{equation}
Since $E(\tau)$ is real, it also has zeros
$\overline{\tau_k}$ in the third quadrant. From (\ref{exp15}) we obtain
that the zeros $\z_q(n)$ of $s_{n-1}(n\z)$ such that $|\z_q(n)-1|\le Mn^{-1/2}$
and $\Im\z_k\ge 0$ are simple and they have the asymptotics
\begin{equation}\label{exp18}
\z_q(n)=1+n^{-1/2}\tau_q+O(n^{-1}),\quad n\to\infty.
\end{equation}
This formula, with the error term $o(n^{-1/2})$, was obtained by Newman
and Rivlin.

{\it Case (2), $M\le |\tau|\le  n^{\frac{1}{6}-\ep}$, $\Re\tau<0$.} 
For large $\tau$ formula (\ref{exp14}) is modified as follows:
\begin{equation}\label{exp18a}
e^{n(1-\z)} \z^n =e^{-\frac{\tau^2}{2}}(1+O(\tau^3n^{-\frac{1}{2}})),
\end{equation}
hence instead of equation (\ref{exp16}) we obtain the equation,
\begin{equation}\label{exp18b}
\frac{1}{\sqrt{2\pi}}
\int_{\tau}^\infty e^{-\frac{x^2}{2}}dx+O(\tau^3n^{-\frac{1}{2}})=0,
\end{equation}
Under the assumption $|\tau|\le  n^{\frac{1}{6}-\ep}$ the error term
is of the order of $O(n^{-3\ep})$. We can rewrite the equation $E(\tau)=0$ in the form,
\begin{equation}\label{exp18c}
\tilde E(\tau)\equiv \frac{1}{\sqrt{2\pi}}
\int_{-\infty}^\tau e^{-\frac{x^2}{2}}dx=1.
\end{equation}
For $\Re\tau\to-\infty$,
\begin{equation}\label{exp18d}
\tilde E(\tau)\sim
-\frac{1}{\sqrt{2\pi}}\tau^{-1}e^{-\frac{\tau^2}{2}},
\end{equation}
hence
\begin{equation}\label{exp18e}
-\frac{1}{\sqrt{2\pi}}\tau_q^{-1}e^{-\frac{\tau_q^2}{2}}\sim 1,
\end{equation}
and
\begin{equation}\label{exp18f}
\tilde E'(\tau_q)=\frac{1}{\sqrt{2\pi}}e^{-\frac{\tau_q^2}{2}}\sim -\tau_q,
\end{equation}
If $\de\equiv|(\tau-\tau_q)\tau_q|\ll |\tau_q|^{-1}$ then
\begin{equation}\label{exp18g}
\tilde E'(\tau)=\frac{1}{\sqrt{2\pi}}e^{-\frac{\tau^2}{2}}
=\tilde E'(\tau_q)(1+O(\de)),
\end{equation}
hence
\begin{equation}\label{exp18h}
|\tilde E(\tau)-1|\ge c|\tau-\tau_q|,\quad |\tau-\tau_q|\le \de;\qquad c>0.
\end{equation}
Equation \ref{exp18b} can be rewritten in the form, 
\begin{equation}\label{exp18i}
\tilde E(\tau)=1+O(\tau^3n^{-\frac{1}{2}}).
\end{equation}
From Proposition \ref{hurwitz} we obtain now that
\begin{equation}\label{exp19}
\z_q(n)=1+n^{-1/2}\tau_q+O(n^{-1}q),\quad q\le n^{\frac{1}{3}-\ep},
\qquad n\to\infty.
\end{equation}
This asymptotics gives an extension of the asymptotics of Newman and Rivlin
to $q\le n^{\frac{1}{3}-\ep}$.  

{\it Case (3), $\Re \tau\le -n^{\ep}$,  $\ep>0$.}
Our calculations in this case are based on the formula,
\begin{equation}\label{exp20}
\sum_{k=0}^\infty e^{-\frac{k^2}{2n}}\z^k
=\frac{1}{1-\z}\left[1+O\left(\frac{1}{n(1-|\z|)^2}\right)\right],
\quad |\z|\le \exp(-n^{-\frac{1}{2}+\ep}).
\end{equation}
To prove this formula, observe that
\begin{equation}\label{exp21}
\sum_{k=\sqrt n}^\infty |\z|^k
=O(n\exp(-n^{\ep})),
\quad n\to\infty,
\end{equation}
and
\begin{equation}\label{exp22}
\sum_{k=0}^{\sqrt n} (1-e^{-\frac{k^2}{2n}})|\z|^k
\le \frac{C_0}{n}\sum_{k=0}^{\sqrt n}k^2|z|^k\le \frac{C_1}{n(1-|\z|)^3}.
\end{equation}
This implies (\ref{exp20}). Similarly, from (\ref{exp6}) we obtain
the estimate,
\begin{equation}\label{exp23}
\sum_{k=0}^\infty \frac{n! n^k\z^k}{(n+k)!}-\sum_{k=0}^\infty e^{-\frac{k^2}{2n}}\z^k
=O\left(\frac{1}{n(1-|\z|)^2}\right),
\quad |\z|\le \exp(-n^{-\frac{1}{2}+\ep}).
\end{equation}
We reduce now equation (\ref{exp7a}) to
\begin{equation}\label{exp24}
\frac{e^{-n\z}n^n \z^n}{n!(1-\z)}\left[1+O\left(\frac{1}{n(1-|\z|)^2}\right)\right]=1,
\qquad |\z|\le \exp(-n^{-\frac{1}{2}+\ep}).
\end{equation}
By applying the Stirling formula and by taking the $n$-th root, we obtain the equation,
\begin{equation}\label{exp25}
\frac{e^{1-\z} \z}{\left[\sqrt{2\pi n}(1-\z)\right]^{\frac{1}{n}}}
\left[1+O\left(\frac{1}{n^2(1-|\z|)^2}\right)\right]=\om_q,
\qquad |\z|\le \exp(-n^{-\frac{1}{2}+\ep}),
\end{equation}
$\om_q=e^{\frac{2\pi q i}{n}}$. As an approximation to this equation, consider
the equation 
\begin{equation}\label{exp26}
h(\z)\equiv e^{1-\z} \z=\om_q,
\qquad |\z|\le \exp(-n^{-\frac{1}{2}+\ep}).
\end{equation}
We have that if $\z=1-t$, then
\begin{equation}\label{exp27}
e^{1-\z} \z=1-\frac{t^2}{2}(1+O(t)),\quad t\to 0,
\end{equation}
hence the solution to the equation $h(\z_q)=\om_q$ has the asymptotics,
\begin{equation}\label{exp28}
\z_q=1+\sqrt{\frac{2q}{n}}e^{\frac{3\pi i}{4}}\left(1+O\left(\frac{q}{n}\right)\right),\qquad
\frac{q}{n}\to 0.
\end{equation}
Also,
\begin{equation}\label{exp29}
|h(\z)-\om_q|\ge c\sqrt{\frac{q}{n}}|\z-\z_q|,\quad |\z-\z_q|\le \frac{z_q}{2};
\qquad c>0.
\end{equation}
As a better approximation to equation (\ref{exp25}), consider
the equation 
\begin{equation}\label{exp30}
h^n(\z)\equiv \frac{e^{1-\z} \z}{\left[\sqrt{2\pi n}(1-\z)\right]^{\frac{1}{n}}}
=\om_q,
\qquad |\z|\le \exp(-n^{-\frac{1}{2}+\ep}).
\end{equation}
Observe that
\begin{equation}\label{exp31}
\left[\sqrt{2\pi n}(1-\z)\right]^{\frac{1}{n}}
=1+O\left(\frac{\ln n}{n}\right),
\qquad |\z|\le \exp(-n^{-\frac{1}{2}+\ep}),
\end{equation}
hence by Proposition \ref{hurwitz}, there exists a zero $\z_q^n$ of $h^n$
such that
\begin{equation}\label{exp32}
\z_q^n=\z_q+O\left(\frac{\ln n}{\sqrt {nq}}\right),
\qquad q\ge n^{\ep}.
\end{equation}
This in turn implies that there is a simple zero $\z_q(n)$ of equation
(\ref{exp25}) such that
\begin{equation}\label{exp33}
\z_q(n)=\z_q^n+O\left(\frac{1}{n^{1/2}q^{3/2}}\right),
\qquad q\ge n^{\ep}.
\end{equation}

The following theorem summarizes the results of this appendix.

\begin{theo} For any $\ep>0$,
the zeros $\z_q(n)$ of $s_{n-1}(n\z)$
have  asymptotics (\ref{exp19})  in the interval $1\le q\le n^{\frac{1}{3}-\ep}$
and asymptotics (\ref{exp33}) in the interval $n^\ep\le q\le \frac{n}{2}$.
\end{theo}

\section {Uniform asymptotics of the function $s_n(n\z)$
in the disk $D(1,\de)$}  
 \label{s_n}

The function
\begin{equation}\label{sn1}
s_n(z)\equiv \sum_{k=0}^n\frac{z^k}{k!}
\end{equation}
solves the equation
\begin{equation}\label{sn2}
s_n'=s_n-\frac{z^n}{n!}\,,
\end{equation}
or
\begin{equation}\label{sn3}
(s_ne^{-z})'=-\frac{e^{-z}z^n}{n!}\,.
\end{equation}
In addition,
\begin{equation}\label{sn4}
\lim_{z\to+\infty}s_n(z)e^{-z}=0,
\end{equation}
hence
\begin{equation}\label{sn5}
s_n(z)=e^z\int_z^{+\infty} \frac{e^{-u}u^ndu}{n!}\,.
\end{equation}
This gives that
\begin{equation}\label{sn6}
s_n(n\z)=\frac{n^{n+1}e^{n(\z-1)}}{n!}\int_{\z}^{+\infty} e^{-n\phi(u)}du,
\end{equation}
where
\begin{equation}\label{sn7}
\phi(\z)=\z-\ln \z-1.
\end{equation}
We will assume that $\ln \z$ is taken on the principal branch, with a
cut on $(-\infty,0]$. Observe that $\z=1$ is a critical point
of $\phi(\z)$ and
\begin{equation}\label{sn8}
\phi(\z)=\frac{(\z-1)^2}{2}-\frac{(\z-1)^3}{3}+\ldots
\end{equation}
Therefore, the function 
\begin{equation}\label{sn9}
\xi(\z)=\sqrt{\phi(\z)}=\sqrt{\z-\ln \z-1}\,
\end{equation}
is analytic in some disk $D(1,\de)$,  $\de>0$, and $\xi$ is the conformal
mapping,
\begin{equation}\label{sn10}
\xi: D(1,\de)\to \Om,
\end{equation}
where $\Om$ is a domain with analytic boundary, $0\in \Om$.
It follows from (\ref{sn7}) that $\xi(\z)$ is analytically continued
to the half-line $(0,\infty)$ and $\xi(0)=-\infty$, $\xi(+\infty)=+\infty$.
From (\ref{sn8}) we have that
\begin{equation}\label{sn11}
\xi(\z)=\frac{\z-1}{\sqrt {2}}-\frac{(\z-1)^2}{6\sqrt{2}}+\frac{(\z-1)^3}{36\sqrt{2}}+\ldots
\end{equation}
For the inverse mapping, $\eta=\xi^{-1}:\Om\to D(1,\de)$, we have that
\begin{equation}\label{sn12}
\z=\eta(\xi)=1+\sqrt 2\,\xi+\frac{2\xi^2}{3}+\frac{\sqrt 2\,\xi^3}{18}+\ldots
\end{equation}
After the substitution $v=\xi(u)$, (\ref{sn6}) becomes
\begin{equation}\label{sn13}
s_n(n\z)=\frac{n^{n+1}e^{n(\z-1)}}{n!}
\int_{\xi(\z)}^{+\infty} e^{-n v^2}\eta'(v)\,dv,
\end{equation}
or if we put $w=\sqrt n \,v$,
\begin{equation}\label{sn14}
s_n(n\z)=\frac{n^{n+\frac{1}{2}}e^{n(\z-1)}}{n!}
\int_{\sqrt n\,\xi(\z)}^{+\infty} e^{-w^2}\eta'\left(\frac{w}{\sqrt n}\right)\,dw.
\end{equation}
By applying the Stirling formula, we obtain that
\begin{equation}\label{sn15}
\frac{s_n(n\z)}{e^{n\z}}=\frac{e^{-\frac{\theta}{12n}}}{\sqrt{2\pi}}
\int_{\sqrt n\,\xi(\z)}^{+\infty} e^{-w^2}\eta'\left(\frac{w}{\sqrt n}\right)\,dw.
\end{equation}
The asymptotics of the integral on the right is described in terms of
the complementary error function,
\begin{equation}\label{sn16}
\Erfc(z)=\frac{2}{\sqrt\pi}\int_z^{+\infty} e^{-w^2}dw.
\end{equation}
For any $\ep>0$, as $|z|\to\infty$,
\begin{equation}\label{sn17}
\Erfc(z)=\frac{e^{-z^2}}{z\sqrt \pi}\left(1-\frac{1}{2z^2}+O(z^{-4})\right),
\quad |\arg z|<\frac{3\pi}{4}-\ep\,,
\end{equation}
see \cite{AS}. Let us fix 
an arbitrary (big) number $M>1$ and consider two cases: 
(1) $|\z-1|\le \frac{M}{\sqrt n}\,$, and 
(2) $\frac{M}{\sqrt n}\le |\z-1|\le \de\,$.

{\it Case 1, $|\z-1|\le \frac{M}{\sqrt n}\,$.} Since
\begin{equation}\label{sn18}
\eta'\left(\frac{w}{\sqrt n}\right)=\eta'(0)+O(n^{-\frac{1}{2}})=\sqrt 2+O(n^{-\frac{1}{2}}),
\end{equation}
if $w$ is bounded, we obtain from (\ref{sn15}) that
\begin{equation}\label{sn19}
\frac{s_n(n\z)}{e^{n\z}}=\frac{1}{2}\Erfc\left(\sqrt n\,\xi(\z)\right)+O(n^{-\frac{1}{2}}),
\quad |\z-1|\le \frac{M}{\sqrt n}\,.
\end{equation}

{\it Case 2, $\frac{M}{\sqrt n}\le |\z-1|\le \de\,$.} Suppose first that 
\begin{equation}\label{sn20}
|\arg (\z-1)|\le \frac{2\pi}{3}\,.
\end{equation}
Then by (\ref{sn11}),
\begin{equation}\label{sn21}
|\arg (\xi(\z))|< 0.7\pi<\frac{3\pi}{4}\,,
\end{equation}
if $\de$ is small enough. Set $a=\sqrt n\,\xi(\z)$. We have that
\begin{equation}\label{sn22}
\begin{aligned}
\int_a^{+\infty} e^{-w^2}\eta'\left(\frac{w}{\sqrt n}\right)\,dw
&=\eta'\left(\frac{a}{\sqrt n}\right)\,\int_a^{+\infty} e^{-w^2}dw\\
&+\int_a^{+\infty} e^{-w^2}
\left[\eta'\left(\frac{w}{\sqrt n}\right)-\eta'\left(\frac{a}{\sqrt n}\right)\right]\,\,dw,
\end{aligned}
\end{equation}
and as $n\to\infty$,
\begin{equation}\label{sn23}
\int_a^{+\infty} e^{-w^2}
\left[\eta'\left(\frac{w}{\sqrt n}\right)-\eta'\left(\frac{a}{\sqrt n}\right)\right]\,\,dw
=O\left(\frac{e^{-a^2}}{a^2\sqrt n}\right)\,,\quad
|\arg a|<\frac{3\pi}{4}-\ep.
\end{equation}
Indeed, set $w=a+t$. Then the latter integral becomes
\begin{equation}\label{sn23a}
\int_0^{+\infty} e^{-a^2-2at-t^2}
\left[\eta'\left(\frac{a+t}{\sqrt n}\right)-\eta'\left(\frac{a}{\sqrt n}\right)\right]\,\,dt.
\end{equation}
We can choose the contour of integration near $t=0$ as $t=re^{-i\arg a}$, $r_0>r>0$,
where $r_0=0.1|a|$,
and then from $t_0=r_0e^{-i\arg a}$ to $+\infty$ in such a way that $|e^{-(a+t)^2}|$ is decreasing to 
0. Observe that for $r_0>r>0$,
\begin{equation}\label{sn23b}
\left|\eta'\left(\frac{a+re^{-i\arg a}}{\sqrt n}\right)-\eta'\left(\frac{a}{\sqrt n}\right)\right|
\le \frac{Cr}{\sqrt n}\,,
\end{equation}
hence 
\begin{equation}\label{sn23c}
\left|\int_0^{t_0} e^{-2|a|r-t^2}
\left[\eta'\left(\frac{a+t}{\sqrt n}\right)-\eta'\left(\frac{a}{\sqrt n}\right)\right]\,\,dt\right|
\le \frac{C}{\sqrt n}\int_0^{r_0}e^{-|a|r}rdr< \frac{C}{\sqrt n\,|a|^2}\,.
\end{equation}
This gives (\ref{sn23}).

Since $\eta(\xi(\z))=\z$ and $a=\sqrt n\,\xi(\z)$, we have that
\begin{equation}\label{sn24}
\eta'\left(\frac{a}{\sqrt n}\right)=\eta'(\xi(\z))=\frac{1}{\xi'(\z)}\,.
\end{equation}
We obtain now from (\ref{sn22}), (\ref{sn23}) and (\ref{sn17}), that
\begin{equation}\label{sn27}
\begin{aligned}
\int_a^{+\infty} e^{-w^2}\eta'\left(\frac{w}{\sqrt n}\right)\,dw
&=\frac{1}{\xi'(\z)}\,\int_a^{+\infty} e^{-w^2}dw
\left[1+O\left(\frac{1}{a\sqrt n}\right)\right]\\
&=\frac{\sqrt{\pi}}{2\xi'(\z)}\,\Erfc(a)
\left[1+O\left(\frac{1}{a\sqrt n}\right)\right]\,,\quad
|\arg a|<\frac{3\pi}{4}-\ep,
\end{aligned}
\end{equation}
hence by (\ref{sn15}),
\begin{equation}\label{sn28}
\frac{s_n(n\z)}{e^{n\z}}=
\frac{1}{2\sqrt{2}\,\xi'(\z)}\,\Erfc(\sqrt{n}\,\xi(\z))
\left[1+
O\left(\frac{1}{(\z-1) n}\right)\right]\,,
\quad |\arg (\z-1)|\le \frac{2\pi}{3}\,.
\end{equation}

Suppose now that 
\begin{equation}\label{sn29}
|\arg (\z-1)-\pi|\le \frac{2\pi}{3}\,.
\end{equation}
Then by (\ref{sn11}),
\begin{equation}\label{sn30}
|\arg (\xi(\z))-\pi|< 0.7\pi<\frac{3\pi}{4}\,,
\end{equation}
if $\de$ is small enough. Observe that $s_n(0)=1$, 
hence from (\ref{sn6}) we obtain that 
\begin{equation}\label{sn31}
\int_0^{+\infty} e^{-n\phi(u)}du=\frac{e^n n!}{n^{n+1}}\,.
\end{equation}
Therefore, (\ref{sn6}) can be rewritten as
\begin{equation}\label{sn32}
s_n(n\z)=e^{n\z}-\frac{n^{n+1}e^{n(\z-1)}}{n!}\int_0^{\z} e^{-n\phi(u)}du,
\end{equation}
and (\ref{sn15}) as
\begin{equation}\label{sn33}
\frac{s_n(n\z)}{e^{n\z}}=1-\frac{e^{-\frac{\theta}{12n}}}{\sqrt{2\pi}}
\int_{-\infty}^{\sqrt n\,\xi(\z)} e^{-w^2}\eta'\left(\frac{w}{\sqrt n}\right)\,dw,
\end{equation}
Observe that
\begin{equation}\label{sn34}
\Erfc(-z)=\frac{2}{\sqrt\pi}\int_{-\infty}^z e^{-w^2}dw,
\end{equation}
and by (\ref{sn17}),
\begin{equation}\label{sn35}
\Erfc(-z)=-\frac{e^{-z^2}}{z\sqrt \pi}\left(1-\frac{1}{2z^2}+O(z^{-4})\right),
\quad |\arg z-\pi|<\frac{3\pi}{4}-\ep\,.
\end{equation}
Also,
\begin{equation}\label{sn36}
\Erfc(-z)+\Erfc z=2.
\end{equation}
Similar to (\ref{sn28}), we obtain now that
\begin{equation}\label{sn37}
\frac{s_n(n\z)}{e^{n\z}}=
1-\frac{1}{2\sqrt{2}\,\xi'(\z)}\,\Erfc(-\sqrt{n}\,\xi(\z))
\left[1+
O\left(\frac{1}{(\z-1) n}\right)\right]\,,
\quad |\arg (\z-1)-\pi|\le \frac{2\pi}{3}\,.
\end{equation}
Let us summarize the results of this appendix.

\begin{theo} \label{wimp}
There exists $\de>0$ such that for any $M>1$
as $n\to\infty$, 
\begin{equation}\label{sn38}
\frac{s_n(n\z)}{e^{n\z}}=\left\{
\begin{aligned}
&\frac{1}{2}\,\Erfc\left(\sqrt n\,\xi(\z)\right)+O(n^{-\frac{1}{2}}),
\quad {\rm if}\quad |\z-1|\le \frac{M}{\sqrt n}\,;\\
&\frac{1}{2\sqrt{2}\,\xi'(\z)}\,\Erfc(\sqrt{n}\,\xi(\z))
\left[1+
O\left(\frac{1}{(\z-1) n}\right)\right]\,,\\
&\hskip 2cm  {\rm if}\quad
\frac{M}{\sqrt n}\le |\z-1|\le \de
\quad {\rm and}\quad |\arg (\z-1)|\le \frac{2\pi}{3}\,;\\
&1-\frac{1}{2\sqrt{2}\,\xi'(\z)}\,\Erfc(-\sqrt{n}\,\xi(\z))
\left[1+
O\left(\frac{1}{(\z-1) n}\right)\right]\,,\\
&\hskip 2cm  {\rm if}\quad
\frac{M}{\sqrt n}\le |\z-1|\le \de
\quad {\rm and}\quad |\arg (\z-1)-\pi|\le \frac{2\pi}{3}\,.
\end{aligned}\right.
\end{equation}
\end{theo}

A similar asymptotics for real $\z>0$ was obtained by Jet Wimp (unpublished),
see \cite{GB}. Asymptotics (\ref{sn38}) can be used to locate the zeros of $s_n$.
The zeros $\sg_q$, $\bar\sg_q$, of $\Erfc(\sg)$ are located in the second and the third quadrants,
and the ones in the second quadrant have the asymptotics \cite{FCC},
\begin{equation}\label{sn39}
\sg_q=\sqrt{2\pi q}e^{\frac{3\pi i}{4}}+
\frac{1}{4\sqrt{2\pi q}}\ln(8\pi q) e^{\frac{\pi i}{4}}+
O\left( q^{-\frac{1}{2}}\right),\qquad
q\to\infty,
\end{equation}
cf. (\ref{exp17a}), where $\tau_q=\sqrt 2\,\sg_q$. The first zero is
\begin{equation}\label{sn40}
\sg_1=-1.3548101281\ldots+i1.9914668430\ldots
\end{equation}
Asymptotics (\ref{sn39})
can be obtained from the equation
\begin{equation}\label{sn41}
2=\Erfc(-\sg_q)=-\frac{e^{-\sg_q^2}}{\sg_q\sqrt \pi}\left(1-\frac{1}{2\sg_q^2}+O(\sg_q^{-4})\right),
\qquad
q\to\infty,
\end{equation}

It follows from (\ref{sn38}) that the zeros $\z_q(n)$, $\overline{\z_q(n)}$
of $s_n(n\z)$ are also located in the second and the third quadrants. 
Let us find the large $n$
asymptotics of $\z_q(n)$. First we consider the problem informally.
By (\ref{sn38}), the equation $s_n(n\z)=0$ can be rewritten as
\begin{equation}\label{sn42}
\Erfc(-\sqrt{n}\,\xi(\z))=\frac{2\sqrt{2}}{\eta'(\xi(\z))}\,
\left[1+
O\left(\frac{1}{(\z-1) n}\right)\right].
\end{equation}
Set
\begin{equation}\label{sn43}
\sg=\sqrt{n}\,\xi(\z);
\end{equation}
then (\ref{sn42}) reads
\begin{equation}\label{sn44}
\Erfc(-\sg)=\frac{2\sqrt{2}}{\eta'\left(\frac{\sg}{\sqrt n}\right)}\,
\left[1+
O\left(\frac{1}{\sg\sqrt n}\right)\right].
\end{equation}
We are looking for $\sg=\sg_q+\tau$, where $\tau$ is a small correction
to be determined. It is convenient to take logarithm of the both sides of (\ref{sn44}),
\begin{equation}\label{sn45}
\log\Erfc(-\sg_q-\tau)=\frac{3\ln 2}{2}
-\log\eta'\left(\frac{\sg_q+\tau}{\sqrt n}\right)\,
+O\left(\frac{1}{\sqrt {nq}}\right)\,.
\end{equation}
(Observe that $\sg_q^{-1}=O(q^{-1/2})$.) We have the Taylor expansion,
\begin{equation}\label{sn46}
\begin{aligned}
\log\Erfc(-\sg_q-\tau)&=\log\Erfc(-\sg_q)-\frac{\Erfc'(-\sg_q)\,\tau}{\Erfc(-\sg_q)}
+O(\tau^2)\\
&=\ln 2+\frac{2e^{-\sg_q^2}\,\tau}{\sqrt\pi\,\Erfc(-\sg_q)}+O(\tau^2),
\end{aligned}
\end{equation}
where the error term is uniform in $\sg_q$.  Indeed, it follows from asymptotics (\ref{sn35})
that
\begin{equation}\label{sn47}
\frac{\Erfc'(-\sg)}{\Erfc(-\sg)}=-\frac{2e^{-\sg^2}}{\sqrt\pi\,\Erfc(-\sg)}
=2\sg-\frac{1}{\sg}+O(\sg^{-3})\,,
\end{equation}
hence
\begin{equation}\label{sn47a}
[\log\Erfc(-\sg)]''=O(1),\quad \sg\to\infty;\qquad |\arg \sg-\pi|<\frac{3\pi}{4}\,.
\end{equation}
We find $\tau$ from the equation,
\begin{equation}\label{sn48}
\ln 2+\frac{2e^{-\sg_q^2}\,\tau}{\sqrt\pi\,\Erfc(-\sg_q)}
=\frac{3\ln 2}{2}-\log\eta'\left(\frac{\sg_q}{\sqrt n}\right)\,,
\end{equation}
so that
\begin{equation}\label{sn49}
\tau=
\frac{\sqrt\pi}{2}\left[\frac{\ln 2}{2}-\log\eta'\left(\frac{\sg_q}{\sqrt n}\right)\right]
e^{\sg_q^2}\Erfc(-\sg_q)\,,
\end{equation}
or
\begin{equation}\label{sn50}
\tau=\frac{1}{\sqrt n}\,g\left(\frac{\sg_q}{\sqrt n}\right)h(\sg_q),
\end{equation}
where the function
\begin{equation}\label{sn51}
g(z)=-\frac{1}{2z}\left[\frac{\ln 2}{2}-\log\eta'\left(z\right)\right]=\frac{\sqrt 2}{3}-\frac{5z}{36}+\ldots
\end{equation}
is analytic at $z=0$, and
\begin{equation}\label{sn52}
h(\sg)=-\sqrt\pi\,\sg e^{\sg^2}\Erfc(-\sg)=1-\frac{1}{2\sg^2}+O(\sg^{-4})\,,
\qquad \sg\to\infty.
\end{equation}

\begin{theo} \label {semiclass}
There exists $\de>0$ such that all the zeros $\z_q(n)$ of $s_n(n\z)$
in the domain $D(1,\de)\cap \{\Im\z>0\}$ have the asymptotics,
\begin{equation}\label{sn53}
\z_q(n)=\eta\left(\frac{\sg_q}{\sqrt n}+\frac{\tau_q}{n}\right)
+O\left(\frac{1}{n\sqrt q}\right),
\quad n\to\infty,
\end{equation}
where $\eta$ is the inverse function of $\xi(\z)=\sqrt{\z-\ln\z-1}\,$, see (\ref{sn12}), 
$\{\sg_q,\;q=1,2,\ldots\}$ are the zeros of $\Erfc(\sg)$ in the upper half-plane, and 
\begin{equation}\label{sn54}
\tau_q=g\left(\frac{\sg_q}{\sqrt n}\right)h(\sg_q),
\end{equation}
where $g$ and $h$ are defined in (\ref{sn51}) and (\ref{sn52}), respectively.
\end{theo}

\begin{proof} 
Existence. Let us write equation (\ref{sn45}) as
\begin{equation}\label{sn55}
\log\Erfc(-\sg_q-\tau)=\frac{3\ln 2}{2}-\log\eta'\left(\frac{\sg_q+\tau}{\sqrt n}\right)\,
+\ep(\tau),\qquad \ep(\tau)=O\left(\frac{1}{\sqrt n}\right)\,,
\end{equation}
or as
\begin{equation}\label{sn56}
f(\tau)\equiv\log\Erfc(-\sg_q-\tau)-\frac{3\ln 2}{2}+\log\eta'\left(\frac{\sg_q+\tau}{\sqrt n}\right)\,-\ep(\tau)=0.
\end{equation}
Let
\begin{equation}\label{sn57}
\tau^0=\frac{1}{\sqrt n}\,g\left(\frac{\sg_q}{\sqrt n}\right)h(\sg_q),
\end{equation}
Then from (\ref{sn48}) we obtain that
\begin{equation}\label{sn58}
f(\tau^0)=O\left(\frac{1}{\sqrt {nq}}\right),
\end{equation}
and from (\ref{sn46}), (\ref{sn47}), that
\begin{equation}\label{sn59}
|f'(\tau^0)|>c>0\,;\qquad
f''(\tau)=O(1),\quad |\tau-\tau^0|\le n^{-\frac{1}{4}}.
\end{equation}
This implies the existence of a zero $\tau^1$ of $f(\tau)$ such that
\begin{equation}\label{sn60}
\tau^1=\tau^0+O\left(\frac{1}{\sqrt {nq}}\right).
\end{equation}
By (\ref{sn43}), this means that there is a zero $\z^1$ of $s_n(n\z)$
such that
\begin{equation}\label{sn61}
\sg_q+\tau^1=\sqrt{n}\,\xi(\z^1),
\end{equation}
hence
\begin{equation}\label{sn62}
\z^1=\eta\left(\frac{\sg_q+\tau^0}{\sqrt n}+O\left(\frac{1}{n\sqrt {q}}\right)\right),
\end{equation}
which implies (\ref{sn53}). The existence is proved.

Uniqueness. From (\ref{sn35}) it follows that any zero $\sg$ of equation
(\ref{sn44}) must be 
in the disk $D(\sg_q,n^{-\frac{1}{3}})$, but by (\ref{sn59}) there is a unique
zero in this disk. This proves the uniqueness. Theorem \ref{semiclass} is proved.
\end{proof}

It follows from (\ref{sn54}), (\ref{sn51}), and (\ref{sn52}), that $\tau_q$
is uniformly bounded, hence equation (\ref{sn53}) can be rewritten
in the form,
\begin{equation}\label{sn63}
\z_q(n)=\eta\left(\frac{\sg_q}{\sqrt n}\right)
+\eta'\left(\frac{\sg_q}{\sqrt n}\right)\frac{\tau_q}{n}
+O\left(\frac{1}{n\sqrt q}\right),
\quad n\to\infty.
\end{equation}
Equation (\ref{sn53}) implies also that
\begin{equation}\label{sn64}
\xi(\z_q(n))=\frac{\sg_q}{\sqrt n}+\frac{\tau_q}{n}
+O\left(\frac{1}{n\sqrt q}\right),
\quad n\to\infty.
\end{equation}
It follows from Theorem \ref{wimp} and \ref{semiclass} that
\begin{equation}\label{sn65}
\left.\left(\frac{s_n(n\z)}{e^{n\z}}\right)'\right|_{\z=\z_q(n)}
=\sqrt{2n}\,\sg_q\left(1+O(q^{-1})\right)\,.
\end{equation}
Indeed, when we differentiate the last formula in (\ref{sn38}), 
we obtain, with the help of (\ref{sn64})  and (\ref{sn41}), that 
\begin{equation}\label{sn66}
\begin{aligned}
\left.\left(\frac{s_n(n\z)}{e^{n\z}}\right)'\right|_{\z=\z_q(n)}
&=\left.\frac{\sqrt{n}}{2\sqrt 2}\,\Erfc'(-\sqrt n\xi(\z))\right|_{\z=\z_q(n)}
+\left.\Erfc(-\sqrt n\xi(\z))\right|_{\z=\z_q(n)}\,O(1)\\
&=-\left.\frac{\sqrt{n}}{\sqrt {2\pi}}\,e^{-n\xi(\z)^2}\right|_{\z=\z_q(n)}
+O(1)=-\frac{\sqrt{n}}{\sqrt {2\pi}}\,e^{-\sg_q^2}+O(1)
\end{aligned}
\end{equation}
which implies (\ref{sn65}), due to (\ref{sn41}).
If we differentiate the last formula in (\ref{sn38}) twice, 
we obtain similarly, that
\begin{equation}\label{sn67}
\left|\left(\frac{s_n(n\z)}{e^{n\z}}\right)''\right|=O(nq),\quad {\rm if}
\quad |\z-\z_q(n)|\le \frac{1}{\sqrt{nq}}\,.
\end{equation}
By combining (\ref{sn65}) and (\ref{sn67}), we obtain the following result.

\begin{prop}
There exists $c>0$ and $N>0$ such that $\forall\,n>N$,
\begin{equation}\label{sn68}
\left|\frac{s_n(n\z)}{e^{n\z}}\right|\ge \sqrt{nq}\,|\z-\z_q(n)|,\quad { if}
\quad |\z-\z_q(n)|\le \frac{c}{\sqrt{nq}}\,.
\end{equation}
\end{prop}

\end{appendix}


\begin{thebibliography}{99}

\bibitem{AS}
M. Abramowitz and I.A. Stegun,
Handbook of mathematical functions, Dover, N.Y. 1972.
\bibitem{BR}
T. Bergkvist and H. Rullg\aa rd,
{\it On polynomial eigenfunctions for a class of differential operators},
Math. Res. Lett. {\bf 9} (2002), 153--171. 
\bibitem{Buc}
J.D. Buckholtz, {\it A characterization of the exponential series}, Amer. Math. Monthly 
{\bf 73}, Part II (1966), 121--123.
\bibitem{CVW}
A.J. Carpenter, R.S. Varga, and J. Waldvogel, {\it Asymptotics for the zeros of the partial sums of $e^z$. I}.,
 Rocky Mountain J. Math, (1991) 99--120.
\bibitem{CG}
B. Conrey and A. Ghosh,
{\it On the zeros of the Taylor polynomials associated with the exponential function, }
Amer. Math. Monthly {\bf 95}, No. 6 (1988), 528-533.
\bibitem{Die}
J. Dieudonn\'e,
{\it Sur les z\'eros des polynomes-sectiones de $e^x$},
Bull. Soc. Math. France {\bf 70} (1935), 333--351.
\bibitem{ESV}
A. Edrei, E.B. Saff, and R.S. Varga, {\it Zeros of sections of power series},
Lecture Notes in Mathematics, No. 1002, Springer-Verlag (1983)
\bibitem{FCC}
H.E. Fettis, J.C. Caslin, and K.R. Cramer,
{\it Complex zeros of the error function and of the complementary error function}.
Math. Comp. {\bf 27} (1973), 401--404.
\bibitem{GB}
W.M.Y. Goh and R. Boyer,
{\it On the zero attractor on the Euler polynomials}. Preprint.
arXiv:math.CO/0409062.
\bibitem{Kap}
M. Kappert,
{\it On the zeros of the partial sums of $\cos(z)$ and $\sin(z)$,} Numer. Math. {\bf 74}, (1996) 397--417.
\bibitem{KM}
A.B.J. Kuijlaars and K.T.-R. McLaughlin, {\it Asymptotic zero behavior 
of Laguerre polynomials with negative parameter.} Constr. Approx. {\bf 20} (2004), 
no. 4, 497--523.
\bibitem{Lev}
B.Ja. Levin,  Distribution of zeros of entire functions.  
Translations of Mathematical Monographs {\bf 5}. 
American Mathematical Society, Providence, R.I., 1980. xii+523 pp.
\bibitem{NR}
D.J. Newman and T.J. Rivlin, {\it The zeros of the partial sums of the exponential function}, 
J. Approx. Theory 5 (1972), 405--412. Correction: J. Approx. Theory 16 (1976), 299--300.
\bibitem{Ost}
I.V. Ostrovskii, {\it On zero distribution of sections and tails of power series}, 
Israel Math. Conf. Proc. {\bf 15} (2001), 297--310. 
\bibitem{PV}
I.E. Pritsker and R.S. Varga, {\it The Szeg\"o curve, zero distibution and 
weighted approximation},
 Trans. Amer. Math. Soc. {\bf 349}, No. 10, (1997) 4085--4105.
\bibitem{SV}
E.B. Saff and R.S. Varga,
{\it Zero-free parabolic regions for sequences of polynomials,}
SIAM J. Math. Anal. {\bf 7} (1976), 344--357.
\bibitem{Sop}
E. Soprunova, {\it Exponential Gelfond-Khovanskii formula in dimension one}. Preprint.
\bibitem{Sze}
G. Szeg\"o, {\it \"Uber eine Eigenschaft der Exponentialreihe,
Sitzungsber.} Berl. Math. Ges., {\bf 23} (1924), 500--64.
\bibitem{Var}
R.S. Varga, {\it Scientific computation on some mathematical conjectures}, 
Approximation Theory V (C.K. Chui, L.L. Schumacher, and J.D. Ward, eds.), 
pp. 191--209, Academic Press, 1986.
\bibitem{VC1}
R.S. Varga and A.J. Carpenter, {\it Zeros of the partial sums of cos(z) and sin(z). I.,} 
Numer. Algorithms, {\bf 25} (2000), 363-375.
\bibitem{VC2}
R.S. Varga and A.J. Carpenter, 
{\it Zeros of the partial sums of cos(z) and sin(z), II.,} Numer. Math. {\bf 90} (2001), 371-400.
\bibitem{Yil}
Y.C. Yildirim,
{\it A sum over the zeros of partial sums of $e^x$},
J. Ramanujan Math. Soc. {\bf 6} (1991), 51--66.
\bibitem{Zem}
S.M. Zemyan, {\it On the zeroes of the N-th partial sum of the exponential series}, 
Amer. Math. Monthly  (2005), 891--905.

\end{thebibliography}
\end{document}